\documentclass[twocolumn]{aastex61}
\usepackage{amsmath,amsthm,amssymb}

\newcommand{\vel}{\boldsymbol{u}}
\newcommand{\bfe}{\boldsymbol{e}}

\newcommand{\bfu}{\boldsymbol{u}}

\newcommand{\drm}{{\textrm{d}}}
\newcommand{\dz}{{\textrm{d}}z}
\newcommand{\bfg}{\mbox{\boldmath$g$}}

\newcommand\aastex{AAS\TeX}

\received{February, 2018}
\revised{}
\accepted{}
\submitjournal{ApJ}

\shorttitle{\aastex\ sample article}
\shortauthors{Kessar et al.}

\begin{document}

\title{Scale selection in the stratified convection of the solar photosphere}

\correspondingauthor{Mouloud Kessar}
\email{m.kessar@leeds.ac.uk}

\author{Mouloud Kessar}
\affil{Department of Applied Mathematics \\ University of Leeds \\
Leeds  LS2 9JT \\
UK}
\nocollaboration

\author{David W. Hughes}
\affiliation{Department of Applied Mathematics \\ University of Leeds \\
Leeds  LS2 9JT \\
UK}
\nocollaboration

\author{Evy Kersal\'e}
\affiliation{Department of Applied Mathematics \\ University of Leeds \\
Leeds  LS2 9JT \\
UK}
\nocollaboration

\author{Krzysztof A. Mizerski }
\affiliation{Department of Magnetism \\
Institute of Geophysics \\
Polish Academy of Sciences \\
ul. Ksiecia Janusza 64 \\
01-452 Warsaw \\
Poland}
\nocollaboration

\author{Steven M. Tobias}
\affiliation{Department of Applied Mathematics \\ University of Leeds \\
Leeds  LS2 9JT \\
UK}
\nocollaboration

%
%



\begin{abstract}
We examine the role of stratification in determining the scale for turbulent anelastic convection. Motivated by the range of scales observed in convection at the solar photosphere, we perform local numerical simulations of convection for a range of density contrasts in large domains. We analyse both the Eulerian and Lagrangian statistics of the convection and demonstrate that increasing the stratification shifts the scale of the most energetic structures in the flow to smaller scales; furthermore, the relative amplitude of vertical to horizontal flows in the convection decreases with increasing stratification. We discuss the implications of our results to the issue of solar mesogranulation.
%

\end{abstract}


\setcounter{table}{0}

\section{Introduction} \label{sec:intro}

Convection in the solar photosphere is characterized by a wide range of scales. The smallest clearly discernible scale is that of the granules; granular convection itself has a range of scales, with a typical granule being $1$Mm in horizontal extent and with a lifetime of a few minutes. The kinetic energy contained in these scales is transferred to smaller scales through a turbulent cascade, until it is dissipated as heat at scales of the order of a millimetre. At a considerably larger scale are the supergranules, with a typical size of $30$Mm and a lifetime of a few days. The existence of granular and supergranular scales is clear from numerous observational studies \citep[e.g.][]{Rieutord_2010, Hathaway_2015}. There is, however, an additional putative intermediate convective scale, known as \textit{mesogranulation}. A mesogranular scale of convection was first reported by \cite{November_1981_183539}, who used time-averaged velocity measurements to identify a convective scale of the order of $4$Mm with a lifetime of about two hours. Intriguingly though, there is no specific signature detectable in the kinetic energy spectra at this scale that would unambiguously identify a true mesogranular scale, distinct from either large granules or small supergranules \citep[see][and references therein]{Hathaway_2000_1005200809766, Rieutord_2010}.  The existence of mesogranulation has thus been the subject of considerable debate over the last few decades. 
   
From a theoretical perspective, several computational studies have addressed the modelling of mesogranulation. The first such studies, which employed the Boussinesq approximation \citep{Cattaneo_2001,Cattaneo_2003}, did indeed observe larger-scale convective structures, which could be identified as mesogranules, together with smaller cells, which could be identified as granules. In order to go beyond the Boussinesq approximation by incorporating the influence of stratification --- a crucial ingredient of the solar photosphere --- the problem of convective cell structure has also been addressed using codes that solve the fully compressible convection equations.   For instance, \cite{Rincon_2005_200400130}, \cite{Bushby_2012_638067} and \cite{Bushby_2014_201322993} investigated fully compressible convection, though for quite small density contrasts. They also observed the emergence of large convective cells, of a similar scale to those seen in the Boussinesq configuration; in these studies, mesogranules were associated with the most energetic scale on the kinetic energy spectrum. From the point of view of terminology, when comparing the results of simulations with solar observations, certain studies \citep[e.g.][]{Rieutord_2010, Hathaway_2015} associate the highly energetic convective cells with granules, whilst others \citep[e.g.][]{Bushby_2012_638067,Bushby_2014_201322993} refer to these as mesogranules. 


   
Although the effects of density stratification are of course included in the equations of fully compressible convection, the numerical constraints involved in accurately tracking sound waves are severe, ensuring that only fairly small density contrasts can be studied. However, in the Sun, the density stratification close to the surface is pronounced, with a change in density of nearly four orders of magnitude over the outer $2\%$ of the Sun \citep{Stix_1989}. Here, therefore, we propose to study the pivotal role of density stratification by considering the problem of thermal convection under the anelastic approximation, an asymptotic reduction of the full governing equations that retains the effects of stratification, but filters out sound waves \citep[see][]{Gough_1969_0469, Lantz_1999}. Such an approach has been employed for a number of years in global spherical simulations of stars and planets \citep[e.g.][]{Glatzmaier_1981_1981, Clune_1999}. However, for local models with Cartesian geometries, the development of this types of codes is, somewhat surprisingly, rather new. A review of the various computational approaches that have been employed to model stellar convection is provided by \cite{Kupka2017}.

   

The outline of the paper is as follows. Section~\ref{sec:num_mod} presents the mathematical formulation of the problem of thermal convection in the anelastic approximation, with a brief description of the numerical approach we have employed (a fuller description can be found in the Appendix).  Section~\ref{sec:num_result} describes the results of the numerical simulations, employing three different  approaches to  investigating the dependence of the convective cell structure on the stratification of the atmosphere. The connection with the long-standing issue of solar mesogranulation is discussed in Section~\ref{sec:Discussion}.

\section{Mathematical Formulation of Anelastic Convection} \label{sec:num_mod}

 

We consider the problem of anelastic convection between two infinite parallel
planes, at $z=0$ (bottom) and $z=d$ (top). This orientation of the $z$-axis,
opposite to that traditionally used for compressible convection, allows for
a formal identification of the anelastic and Boussinesq equations. As
discussed in the Introduction, several different formulations of the anelastic
approximation can be found in the literature. Here we follow that introduced
by \cite{Lantz_1999}; its results have been compared with those of fully
compressible codes, both for linear studies \citep{Berkoff_2010_521747} and
fully nonlinear simulations \citep{Verhoeven_2015}.

The starting point for the anelastic approximation is to decompose the density
$\rho$, temperature $T$, pressure $p$ and entropy $s$ into an adiabatic
reference state, indicated by overbars, and a perturbation to this state,
indicated by subscripts ``1'':
\begin{align}
\nonumber
&\rho = \rho_r\left (\bar{\rho} + \varepsilon \rho_1 \right), \qquad
T = T_r \left(\bar{T} + \varepsilon T_1 \right), \\
&p = p_r \left(\bar{p} + \varepsilon p_1 \right), \qquad
s = s_r + c_p \varepsilon \left(\bar{s} + s_1 \right),
\end{align}
where $c_p$ is the specific heat at constant pressure, and $\rho_r$, $T_r$,
$p_r$ and $s_r$ are representative values of the density, temperature, pressure and
entropy evaluated at the bottom of the layer. The asymptotically small parameter $\varepsilon$
 is a dimensionless measure of departure from adiabaticity, expressed as
\begin{equation}
\varepsilon = -\frac{d}{T_r}\left( \frac{\drm \bar{T}}{\dz} +\frac{g}{c_p}
\right),
\end{equation}
where $\bfg = -g \, \bfe_z$ is the gravity vector.

The reference state depends only on the height $z$, and takes the form of a
polytrope:
\begin{equation}
  \bar{\rho} = \left(1+\theta z\right)^m, \quad 
  \bar{T} = 1 + \theta z, \quad 
  \bar{s} = \frac{1}{|\theta|} \ln (1+\theta z),
\end{equation}
where $\theta<0$ is the dimensionless temperature jump across the layer and
$m=3/2$ is the adiabatic polytropic index.

In this formulation of the anelastic approximation, it is assumed that the effect of the molecular transport of heat and momentum is much smaller than that induced by turbulent motions. Hence we introduce a turbulent thermal diffusivity, $\kappa$, and a turbulent kinematic viscosity, $\nu$, with representative values at $z=0$, $\kappa_r$ and $\nu_r$ respectively, together with an entropy based diffusion \citep[see][]{ Braginsky_1995_03091929508228992}. Furthermore, we assume that the dynamic viscosity, $\mu = \bar{\rho} \, \nu$, and the
thermal conductivity, $k= \bar{\rho} \,c_p \,\kappa$, are constant (and hence
$\nu$ and $\kappa$ vary with depth).

On scaling lengths with layer depth, $d$, and times with the thermal
relaxation time, $d^2/\kappa_r$, the evolution of the perturbations to the
reference state in velocity, $\bfu$, and entropy, $s$, (dropping subscripts ``1'') is governed by the following
dimensionless set of equations \citep[e.g.][]{Krzysztof_2011_521748}:
\begin{align}
\nonumber
\frac{\partial \vel}{\partial t} + \vel \cdot \nabla \vel = &-\nabla \left( \frac{p}{\bar{\rho}}\right) + Ra Pr \, s \, \bfe_z \\
& \quad \quad + \frac{Pr}{\bar{\rho}} \left( \nabla^2 \vel + \frac{1}{3} \nabla \left( \nabla \cdot \vel \right)  \right) ,  
\label{eq:mom}
\end{align}
\begin{equation}
\label{eq:mass}
\nabla.\left(\bar{\rho} \vel\right)  = 0 ,
\end{equation}
\begin{equation}
\label{eq:entropy}
\frac{\partial s}{\partial t} + \vel \cdot \nabla s = \frac{ u_z}{1+\theta z}
+ \frac{1}{\bar{\rho}} 
\left( \nabla^2 s +\frac{ \theta}{ \bar{T} } \frac{\partial s}{\partial z} \right)
- \frac{\theta \,Q}{Ra \bar{T}\bar{\rho}},
\end{equation}
\begin{multline}
  \label{eq:Q}
  \text{with} \quad Q =   2 \sum\limits_{i=1}^3 \left(\frac{\partial u_i}{\partial x_i}\right)^2
  + \frac{2}{3}\left(\nabla \cdot \vel \right)^2 \\
  + \sum\limits_{i<j}^3 \left(\frac{\partial u_i}{\partial x_j} +
    \frac{\partial u_j}{\partial x_i} \right)^2. 
\end{multline}

The Rayleigh number, $Ra$, and Prandtl number, $Pr$, are defined by:
\begin{equation}
  Ra = \frac{g  d^3 \varepsilon}{\nu_r \kappa_r}, 
  \qquad
  Pr=\frac{\nu}{\kappa};
\end{equation}
thus $Ra$ is defined by its value at the bottom of the layer, whereas $Pr$ is uniform throughout the layer.
An important point to note is that, with this formulation of the anelastic
equations, the Boussinesq equations are recovered exactly by imposing
$\theta=0$; in this case the variable $s$ is identified not with the entropy,
but with the temperature.

We consider a domain that is square and periodic in
the horizontal directions, of size $\lambda \times \lambda \times 1$. On $z=0$ and $z=1$,
we adopt stress-free and impermeable velocity boundary conditions; the system
is also assumed to have constant entropy (with $s=0$) on $z=0$ and $z=1$. The
layer of fluid is initially at rest with only small random entropy
perturbations.

We have developed a computational code to solve the anelastic
equations~\eqref{eq:mom}--\eqref{eq:entropy} in a Cartesian domain.
Derivatives in the horizontal directions are computed using FFTs, via the FFTW
library, and in the vertical direction by a 4th order finite-difference
representation. Time stepping is achieved through a semi-implicit scheme, in
which the nonlinear terms are treated by a second order Adams-Bashforth method
and the linear terms by a Crank-Nicolson scheme. The pressure is handled via a
poloidal-toroidal decomposition. Implementation of the boundary conditions
requires use of the influence matrix method
\citep{Boronski:2007:PDF:1322570.1322699}. The code has been parallelized
using MPI, with a pencil-based decomposition. Further details of the code are
contained in the Appendix.

\section{Stratified Anelastic Convection: Numerical Results} \label{sec:num_result}

In this section, we explore the influence of density stratification on thermal convection by considering three representative cases. As our benchmark example, we consider Boussinesq (unstratified) convection (i.e.\ $\theta=0$), for which a direct comparison can be made with the results of \cite{Cattaneo_1999}. We then also consider two anelastic configurations (AC) with non-zero (negative) values of $\theta$: a mildly stratified case with $\theta=-0.6$, giving a density contrast $\chi$ across the layer of $\chi = 4$ (AC4), and a strongly stratified case with $\theta=-0.92$ and $\chi= 50$ (AC50). Following \cite{Cattaneo_1999}, for the BC simulation we set $Ra=500\,000$, $Pr=1$ and $\lambda=10$; if $Ra_c$ denotes the critical Rayleigh number for the onset of convection, then the degree of supercriticality is given here by $Ra/Ra_c \approx 760$. As the stratification is increased, it becomes harder to drive convection, i.e.\ $Ra_c$ increases \citep[see e.g.][]{ct:2016}. Thus, in order to make meaningful comparisons between the three cases, we increase $Ra$ for the AC runs in order to maintain the same degree of supercriticality at reference level $z=0$; the parameter values are summarized in Table~\ref{tab:table_parameters}.

We consider the formation and evolution of the convective network from three different perspectives. In Section~\ref{subsec:Convective_net} we describe the broad features of the convection for the three cases by considering the distribution of the temperature (for the Boussinesq case) or the entropy (for the anelastic cases); in Section~\ref{subsec:ink} we look in detail at how the network evolves in time by calculating the dispersion of a passive scalar introduced into the flow; in Section~\ref{subsec:spectra} we analyse the time-averaged convective network by considering the spectral distribution of kinetic energy. Rather than discuss the convection in terms of granules or mesogranules, in these sections we shall refer to either convective cells or the convective network. 

\begin{deluxetable}{ccCrl}[b!]
\tablecaption{Parameter values for the three cases \label{tab:table_parameters}}
\tablecolumns{4}
\tablenum{2}
\tablewidth{0pt}
\tablehead{
\colhead{Parameters} &
\colhead{BC} &
\colhead{AC4} &
\colhead{AC50}  }
\startdata
$\lambda$ & 10 & 10  & 10  \\
$Pr$ & 1 & 1 & 1  \\
$\theta$ & 0 & -0.60 & -0.92  \\
$\chi$ & 1 & 4 & 50  \\
$Ra$ & $5 \times 10^5$ & $1.2 \times 10^6$ & $2.95 \times 10^6$ \\
$Ra_c$ & 657.51 & 1566.89 & 3885.07  \\
\enddata
\end{deluxetable}

\subsection{Convective Networks} \label{subsec:Convective_net}

\begin{figure}[ht!]
\epsscale{1.20}
\plotone{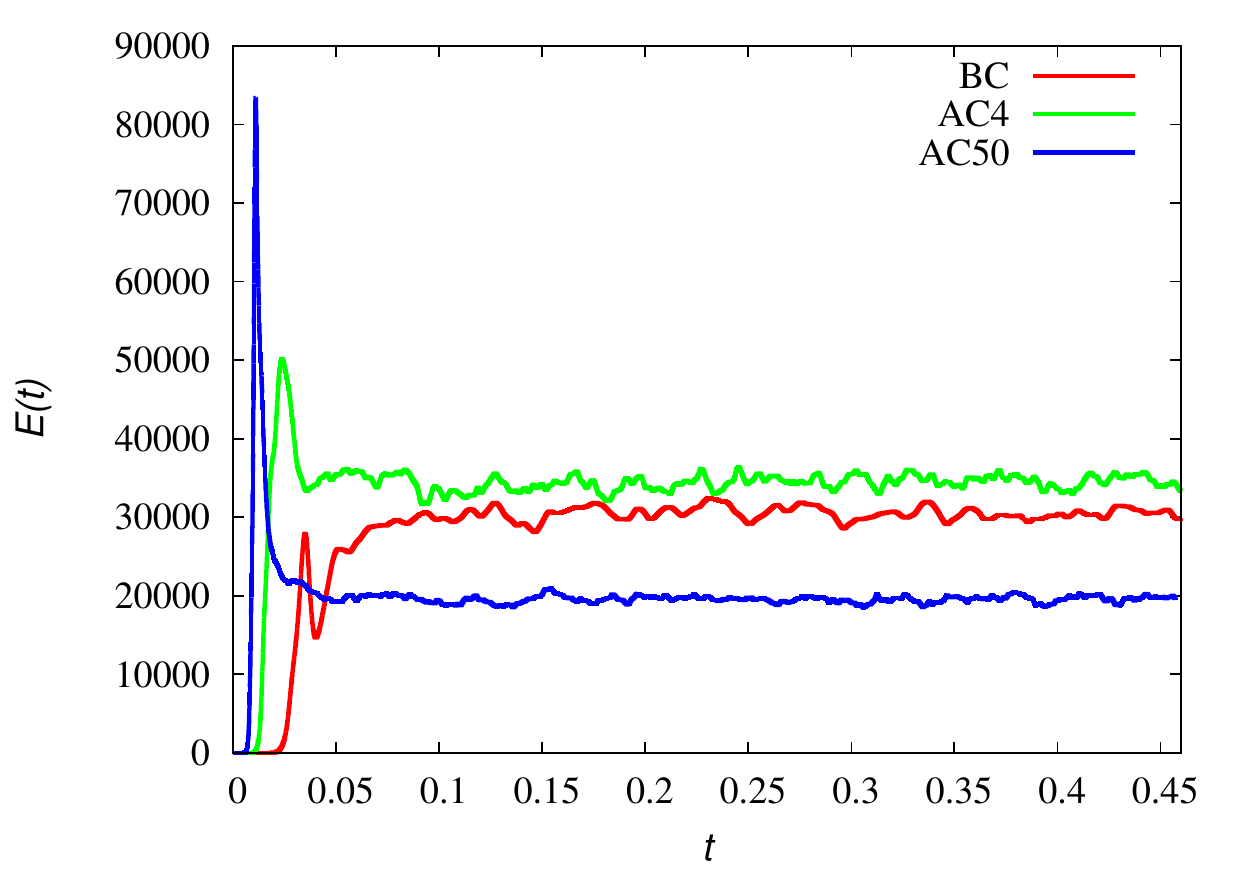}
\caption{Time evolution of the averaged kinetic energy $E(t)$ for the three configurations, with aspect ratio $\lambda = 10$. 
\label{time_evo}}
\end{figure}

Figure~\ref{time_evo} shows the temporal evolution of the volume-averaged
kinetic energy $E$ for the three cases. After an initial (linear) phase of
exponential growth, the energy settles into a statistically stationary state.
It is interesting to note that although the degree of supercriticality (i.e.\
$Ra/Ra_c$) is the same for all three cases, there is some variation in the
level of the energy in the saturated state and, furthermore, that the level of
$E$ is non-monotonic in $\theta$.

\begin{figure*}
\gridline{\fig{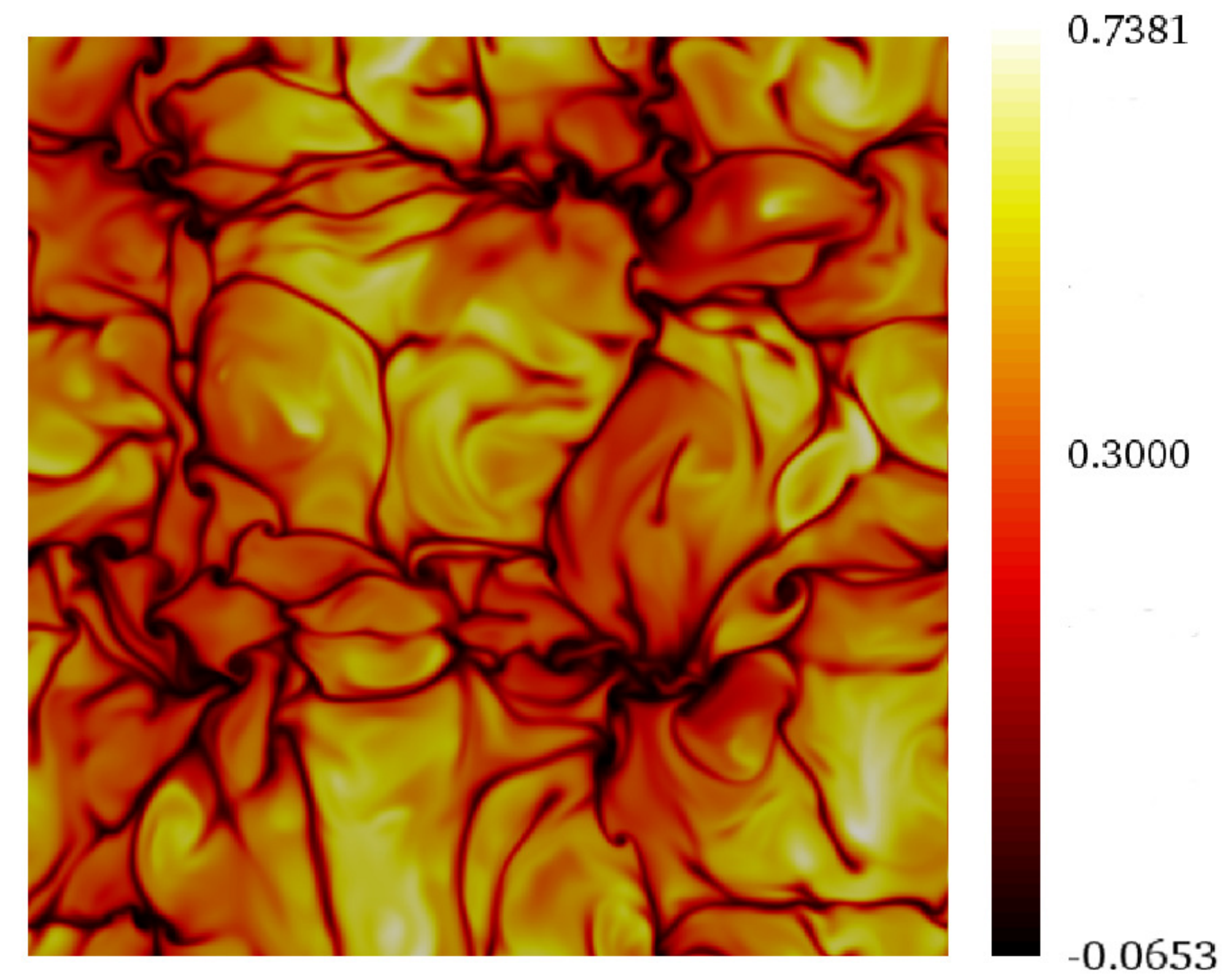}{0.45\textwidth}{}
          \fig{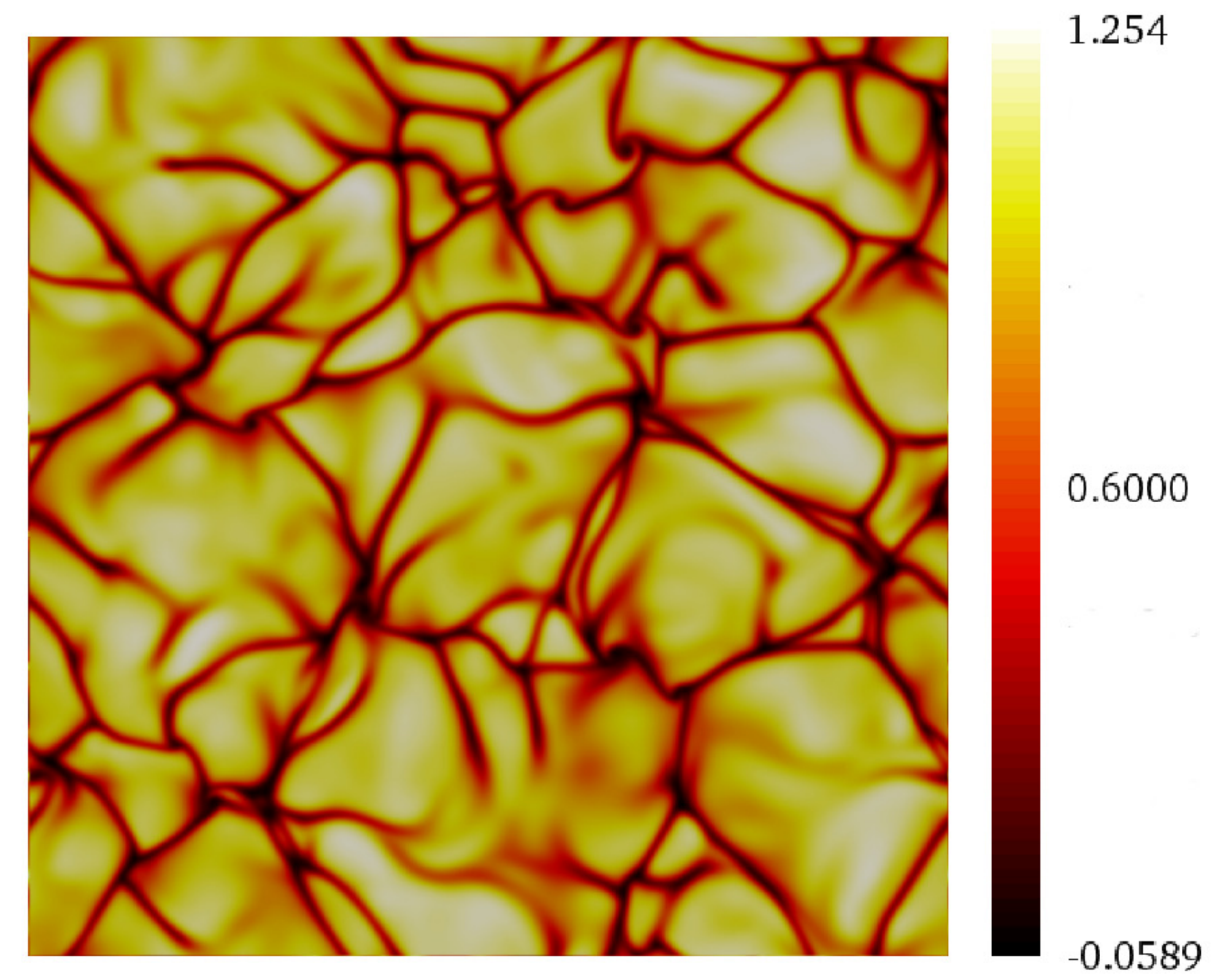}{0.45\textwidth}{}
          }
\gridline{\fig{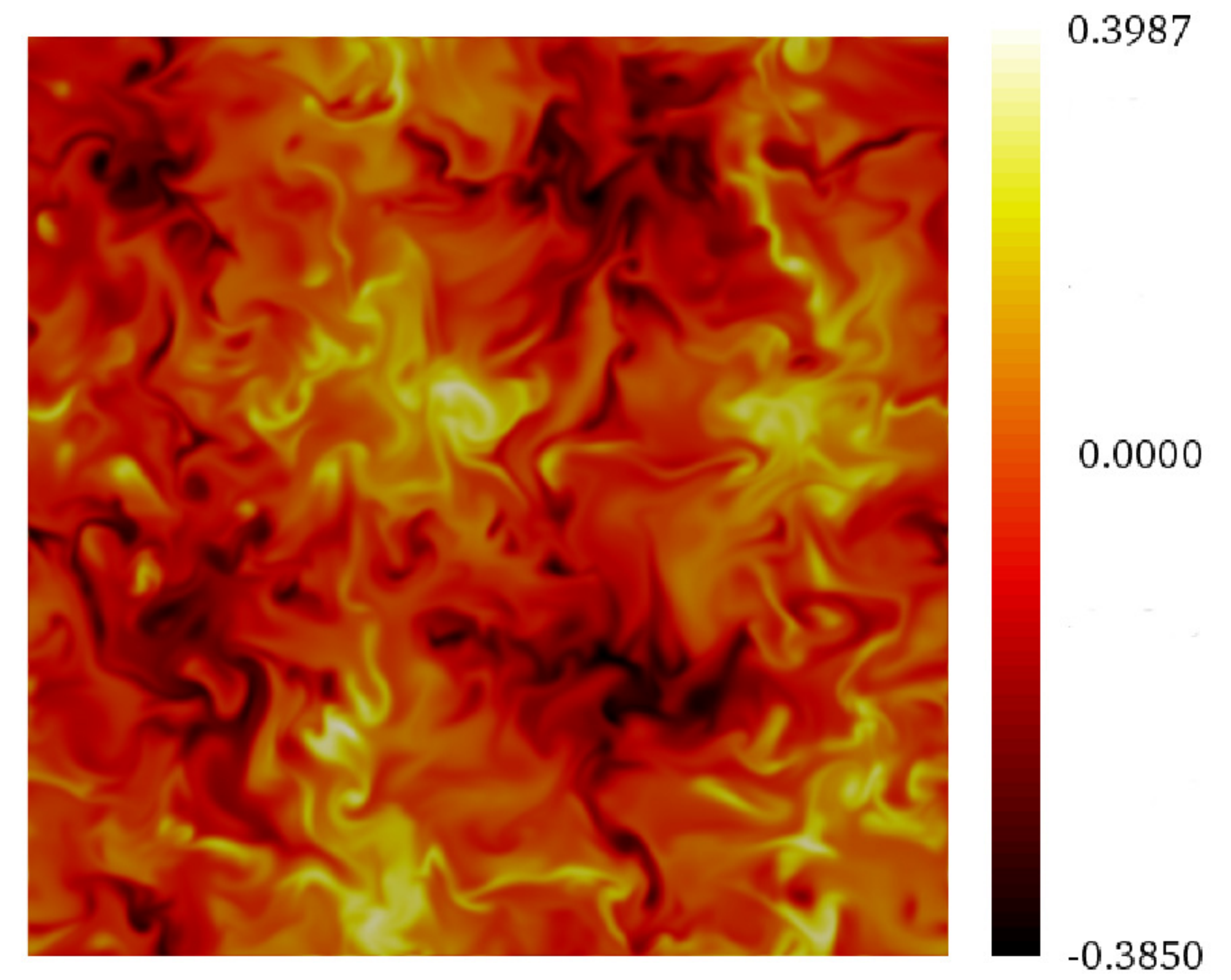}{0.45\textwidth}{}
          \fig{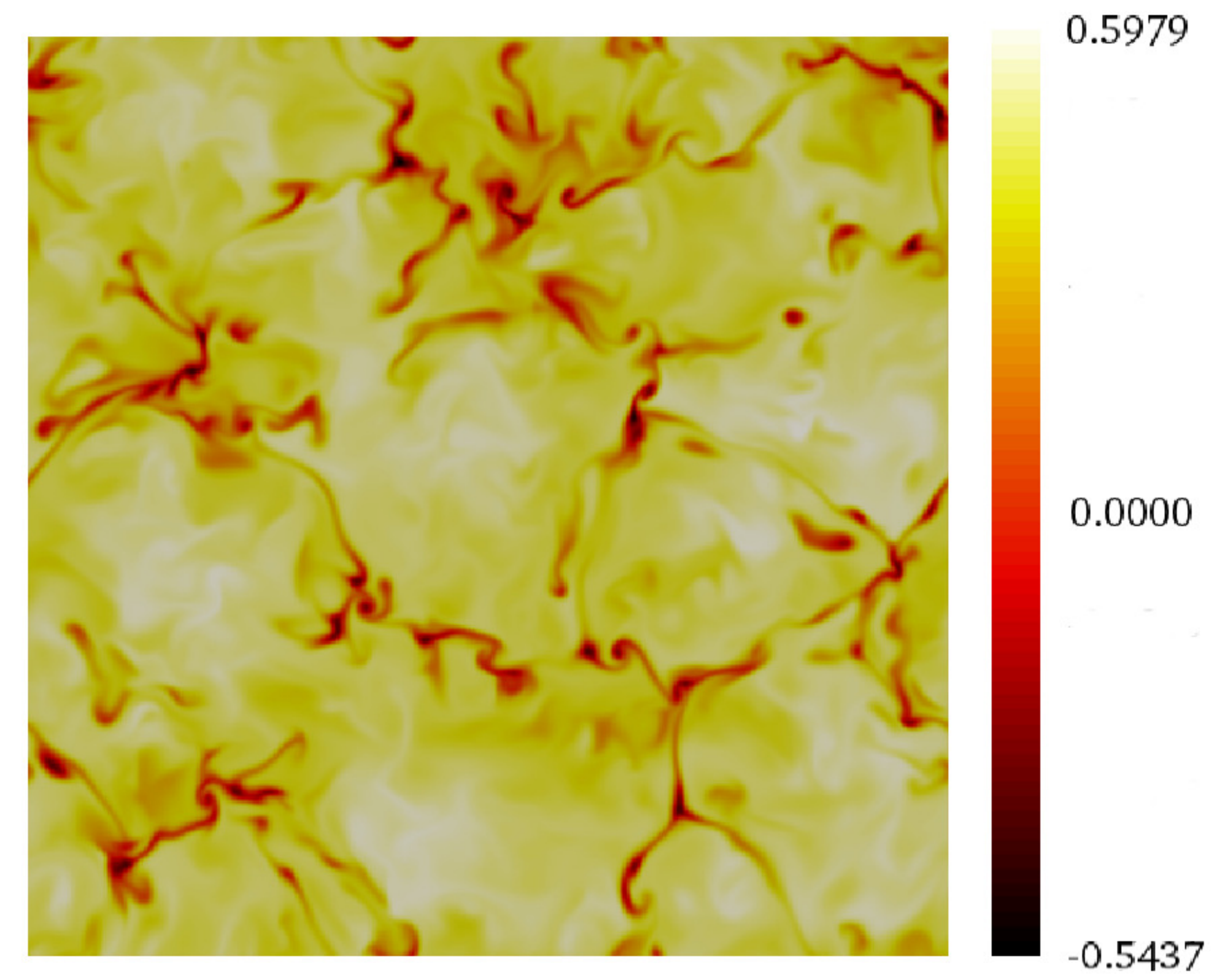}{0.45\textwidth}{}
          }          
\gridline{\fig{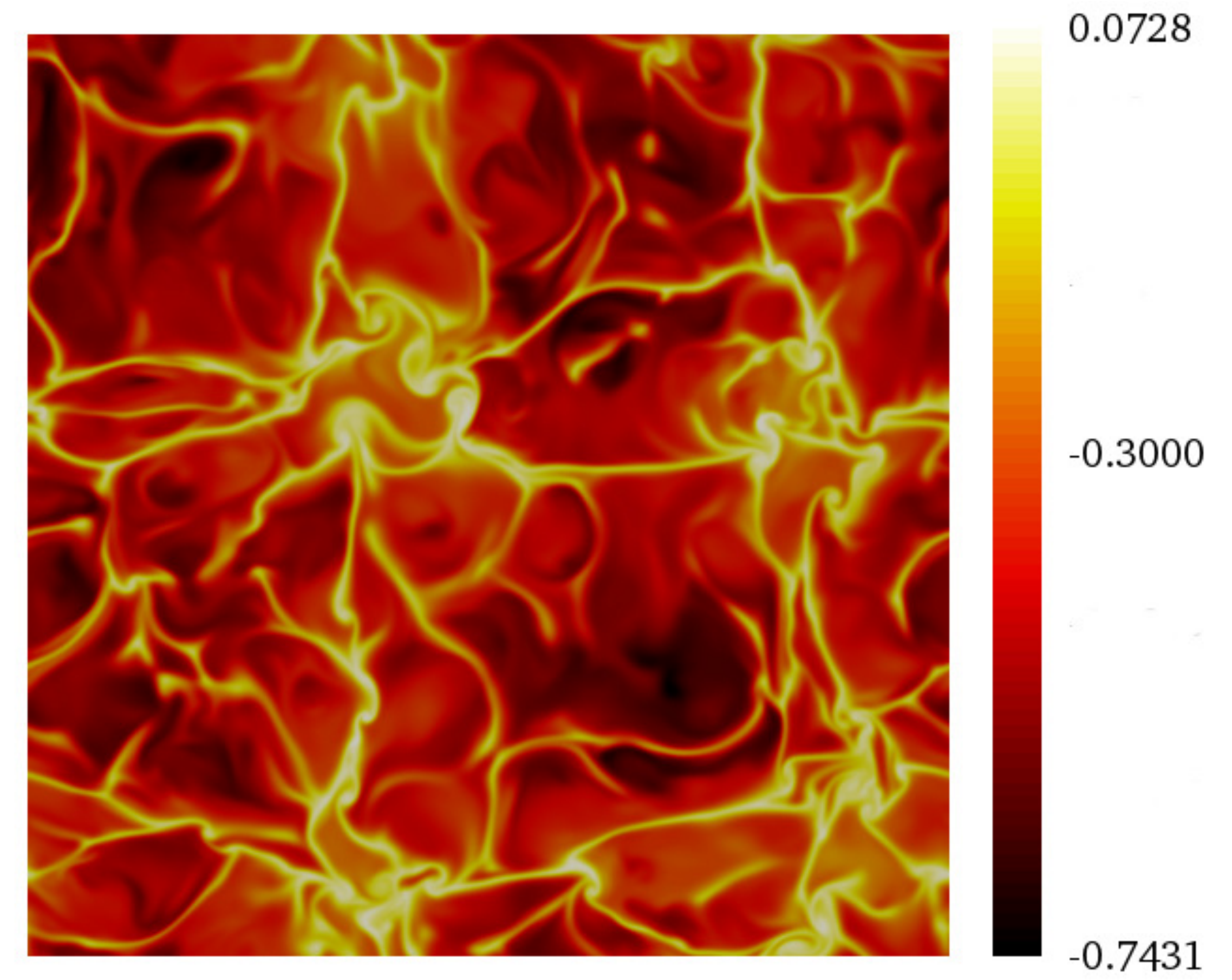}{0.45\textwidth}{(a)}
          \fig{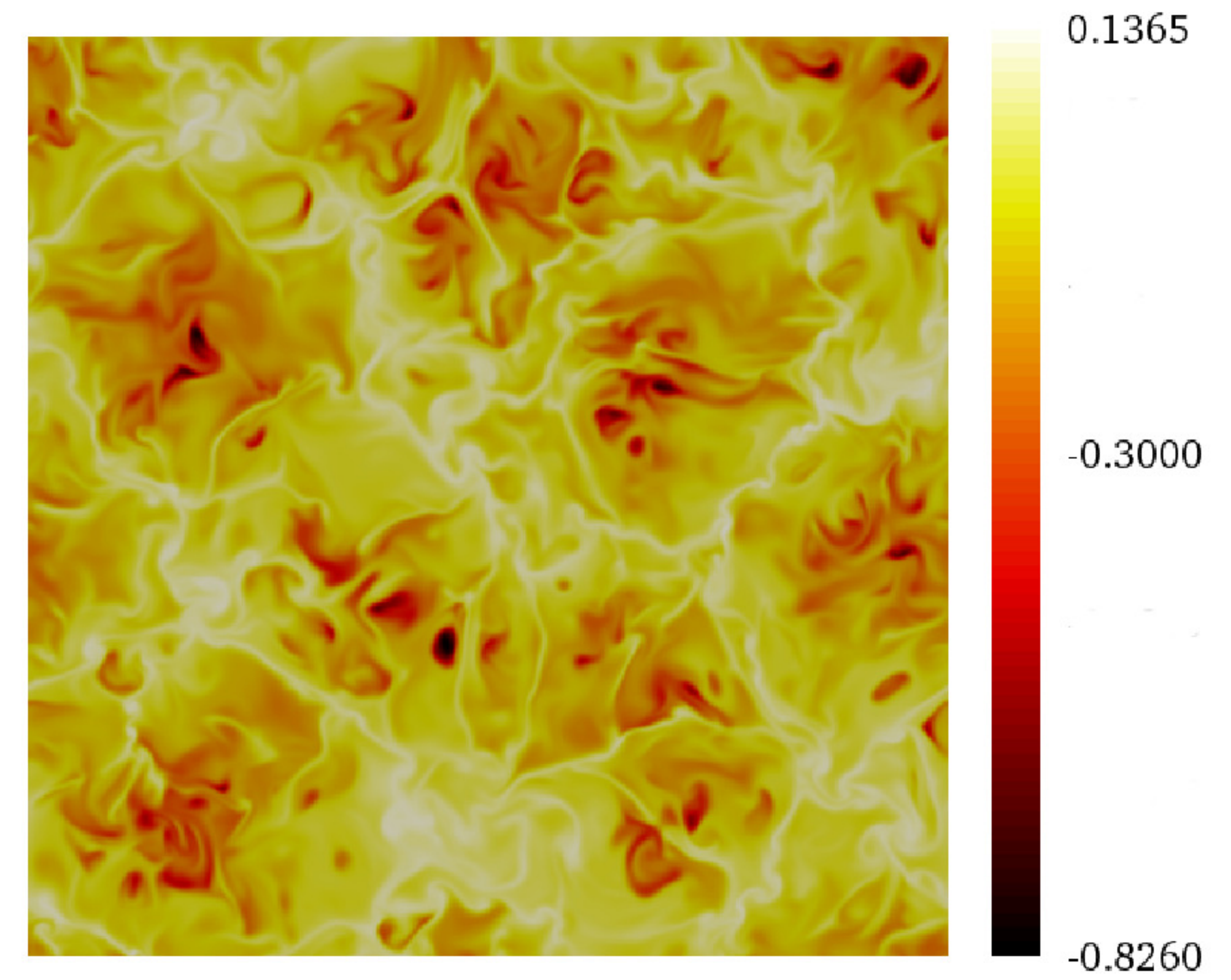}{0.45\textwidth}{(b)}
          }\caption{Horizontal slices of (a)~temperature fluctuations for BC, and (b)~entropy fluctuations for AC4, at heights $z=0.9$ (top), $z=0.5$ (middle) and $z=0.1$ (bottom). \label{Horizontal_slices_BC_AC_4_ratio_10}}
\end{figure*}

The influence of stratification can be seen clearly in
Figure~\ref{Horizontal_slices_BC_AC_4_ratio_10}, which depicts horizontal
slices of temperature or entropy fluctuations at different depths for cases BC
and AC4. Boussinesq convection is characterized by a symmetry about the
mid-plane ($z=0.5$), as can be seen from the general morphology of the
convective networks towards the bottom and top of the domain. Near the bottom
($z=0.1$), one can see a network of convective cells of different sizes,
together with turbulent motions; the large convective cells are associated
with warm upflows. Near the top of the domain ($z=0.9$), the network of cells
is similar to that at the bottom, but now with convective cells delineated by cold
downflows. As expected from the Boussinesq symmetry, the BC case does not
exhibit any particularly distinct cells or structures on the mid-plane. For
anelastic convection, the $z$-symmetry is broken, as can be seen clearly for
the case of AC4 shown in Figure~\ref{Horizontal_slices_BC_AC_4_ratio_10}b.
At the top of the domain, there is a convective network with a range of
scales, not dissimilar to that of Boussinesq convection. On the mid-plane, the
outline of some of the largest cells are still discernible; as previously
observed by \cite{Bushby_2012_638067}, only the strongest downflows penetrate
deeply into the convective region. At the bottom of the domain, and in sharp
contrast to the Boussinesq case, there is no vestige of the convective
network; the flow is highly turbulent, interspersed with light bridges
depicting warm upflows.

\begin{figure}
\gridline{\fig{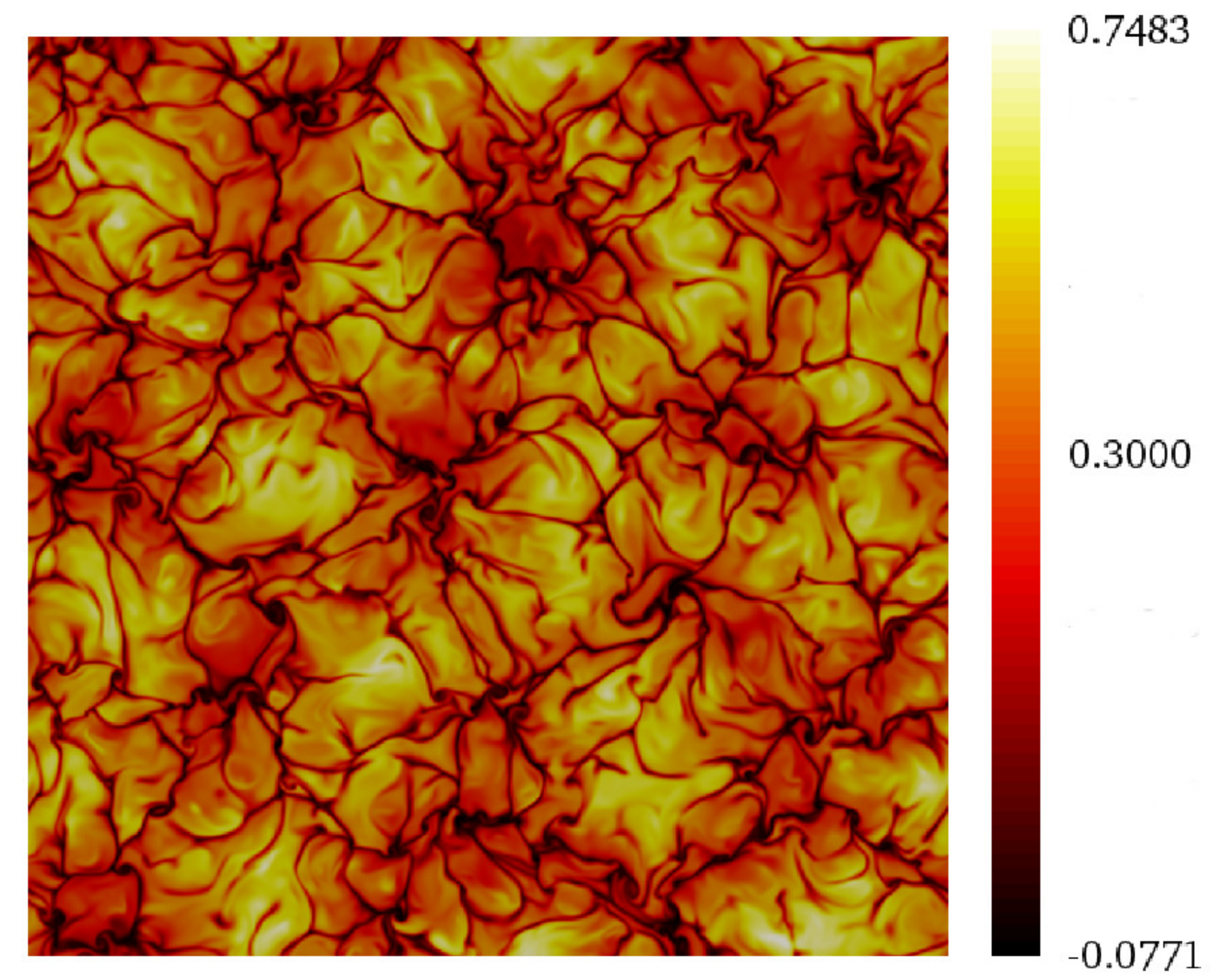}{0.45\textwidth}{(a)}}
\gridline{\fig{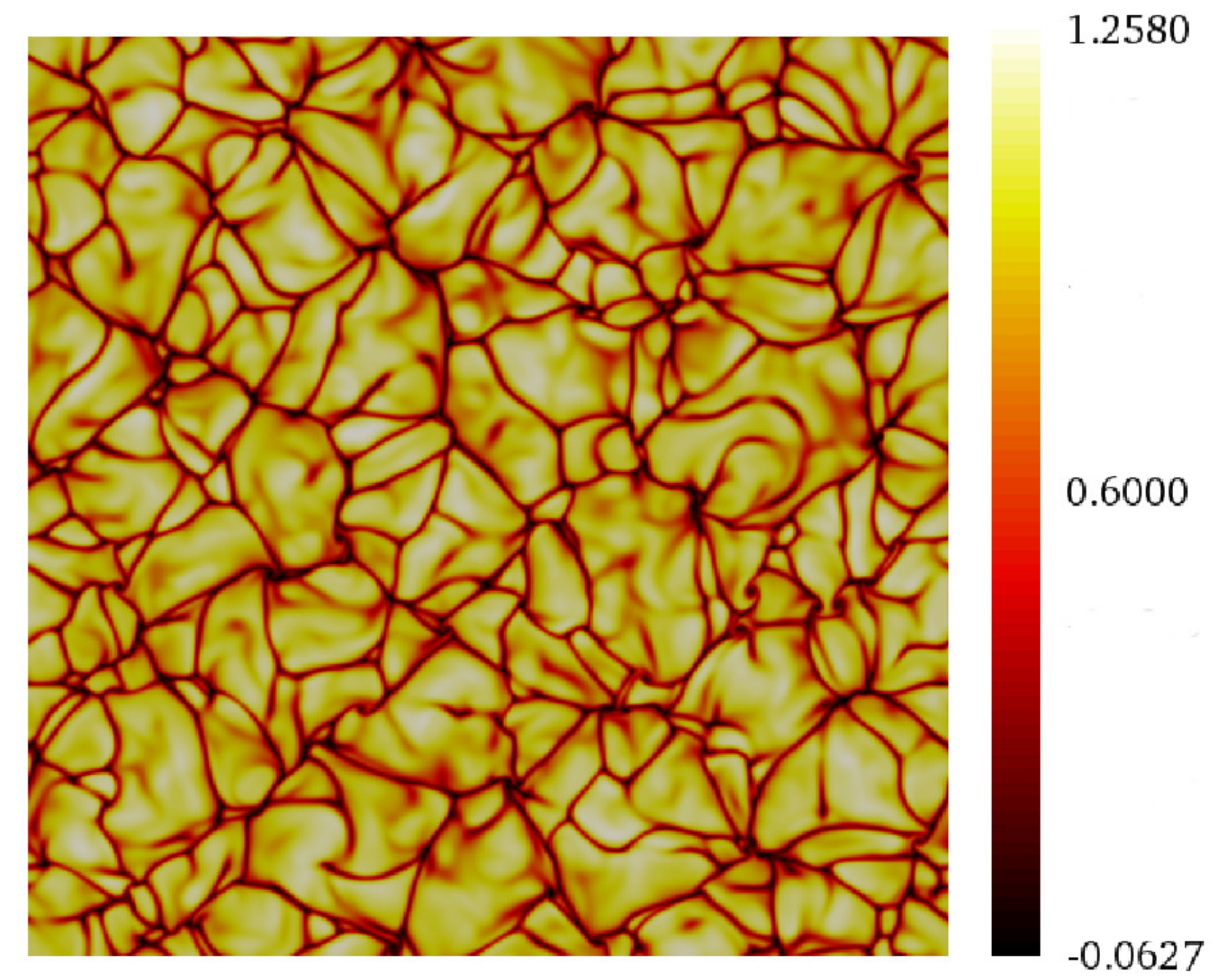}{0.45\textwidth}{(b)}}
          \caption{Horizontal slices of (a) temperature fluctuations for BC, and (b)  entropy fluctuations for AC4; in both cases, the height $z=0.9$ and the aspect ratio $\lambda=20$. \label{Horizontal_slices_BC_AC_4_ratio_20}}
\end{figure}

In order to explore any influence of the horizontal extent of the domain, we have also performed the BC and AC4 simulations at the larger aspect ratio of $\lambda=20$. Figure~\ref{Horizontal_slices_BC_AC_4_ratio_20} shows horizontal slices at $z=0.9$ of temperature (BC) and entropy (AC4) for these two additional configurations, which should be compared with the corresponding cases for $\lambda=10$, shown in the top row of Figure~\ref{Horizontal_slices_BC_AC_4_ratio_10}. For both configurations, the general structure and characteristic sizes are very similar for the $\lambda=10$ and $\lambda=20$ cases, suggesting that with $\lambda=10$ the convection is not constrained by the size of the domain. It should though be pointed out that in the BC case at the larger aspect ratio, the largest convective cells can have more degrees of freedom for their orientation. Thus, for Boussinesq convection, although the essential physics is captured by the $\lambda=10$ configuration, the chaotic temporal evolution of the convective pattern may be slightly constrained by the size of the box.

\begin{figure*}
\gridline{\fig{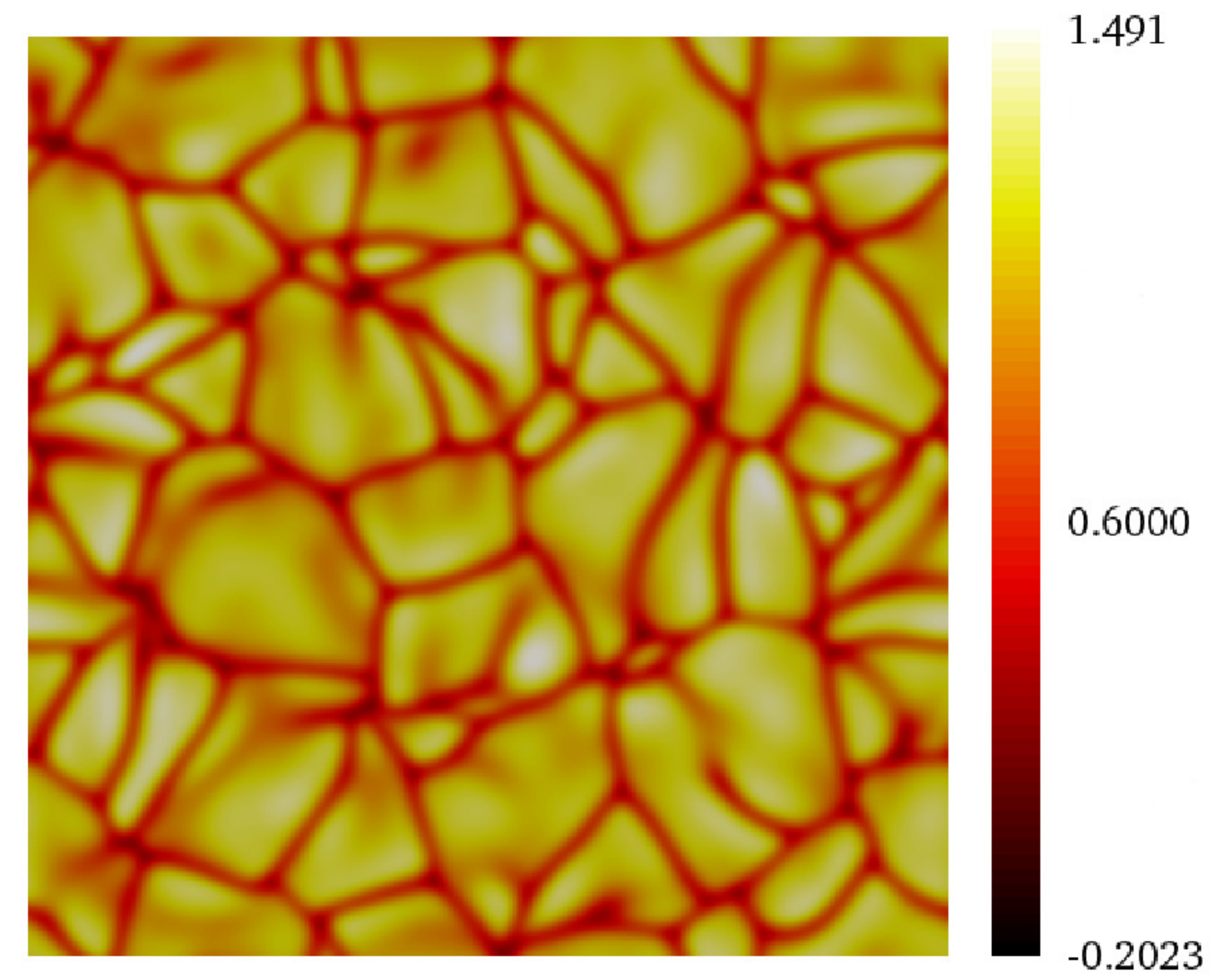}{0.45\textwidth}{}
          \fig{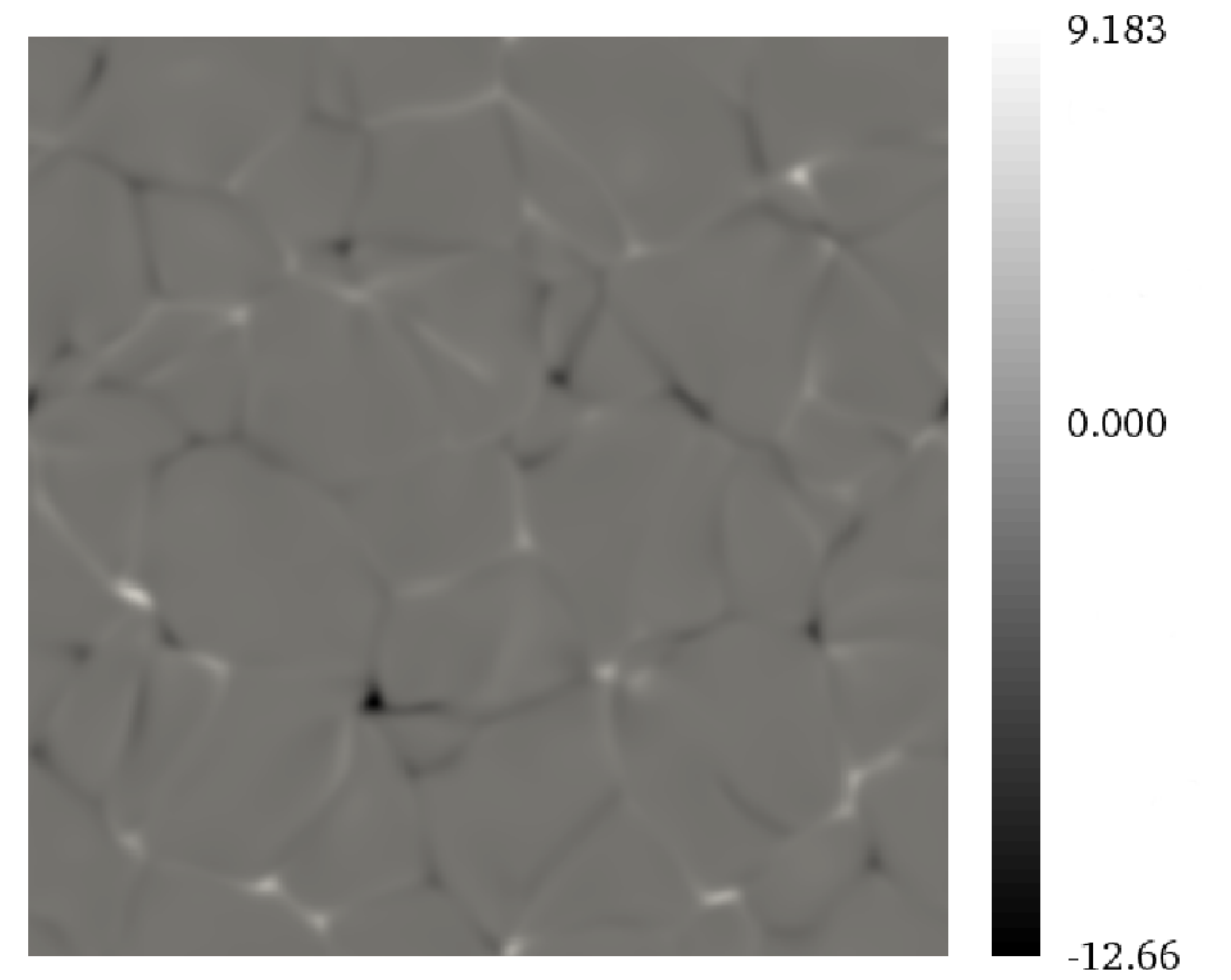}{0.45\textwidth}{}
          }
\gridline{\fig{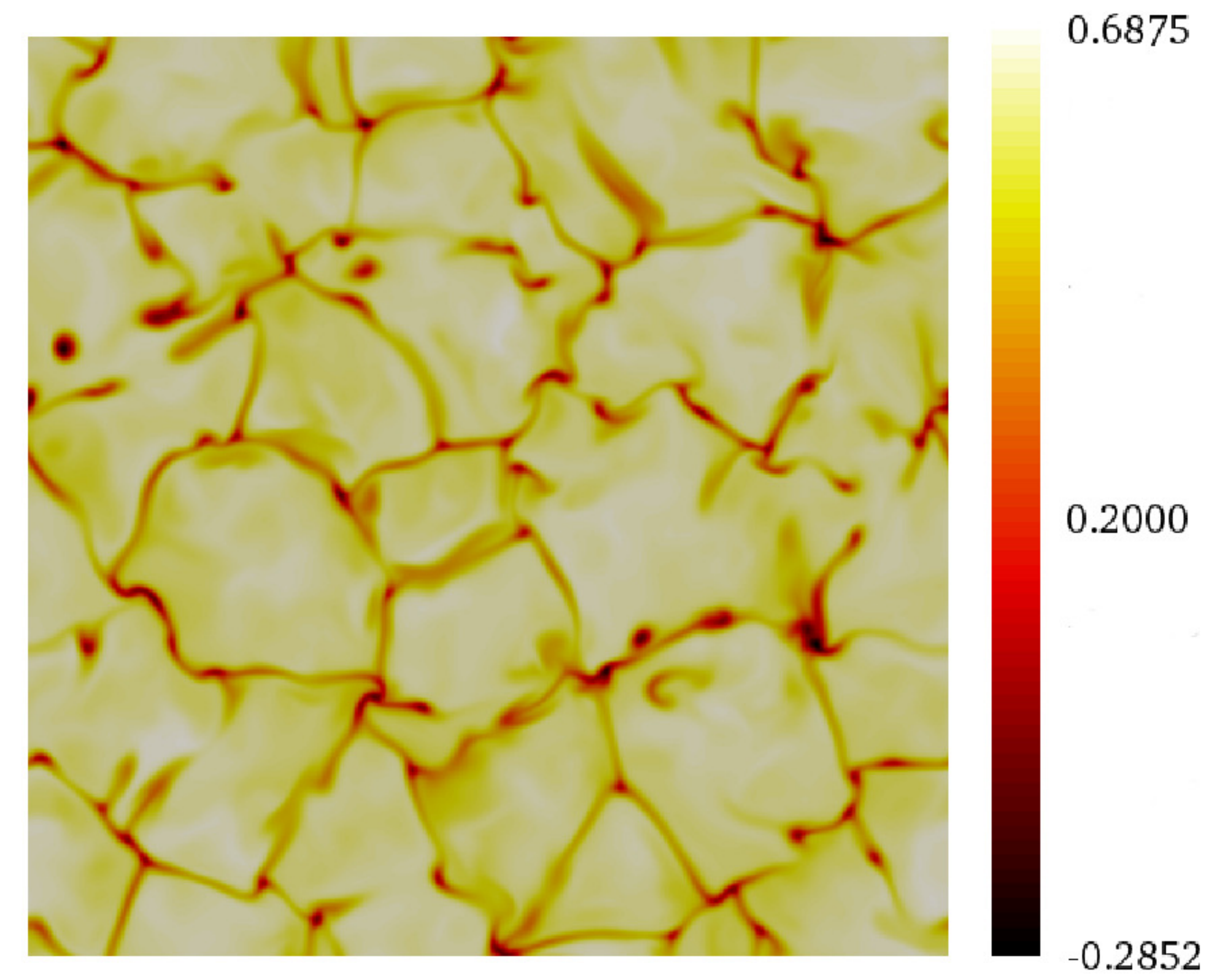}{0.45\textwidth}{}
          \fig{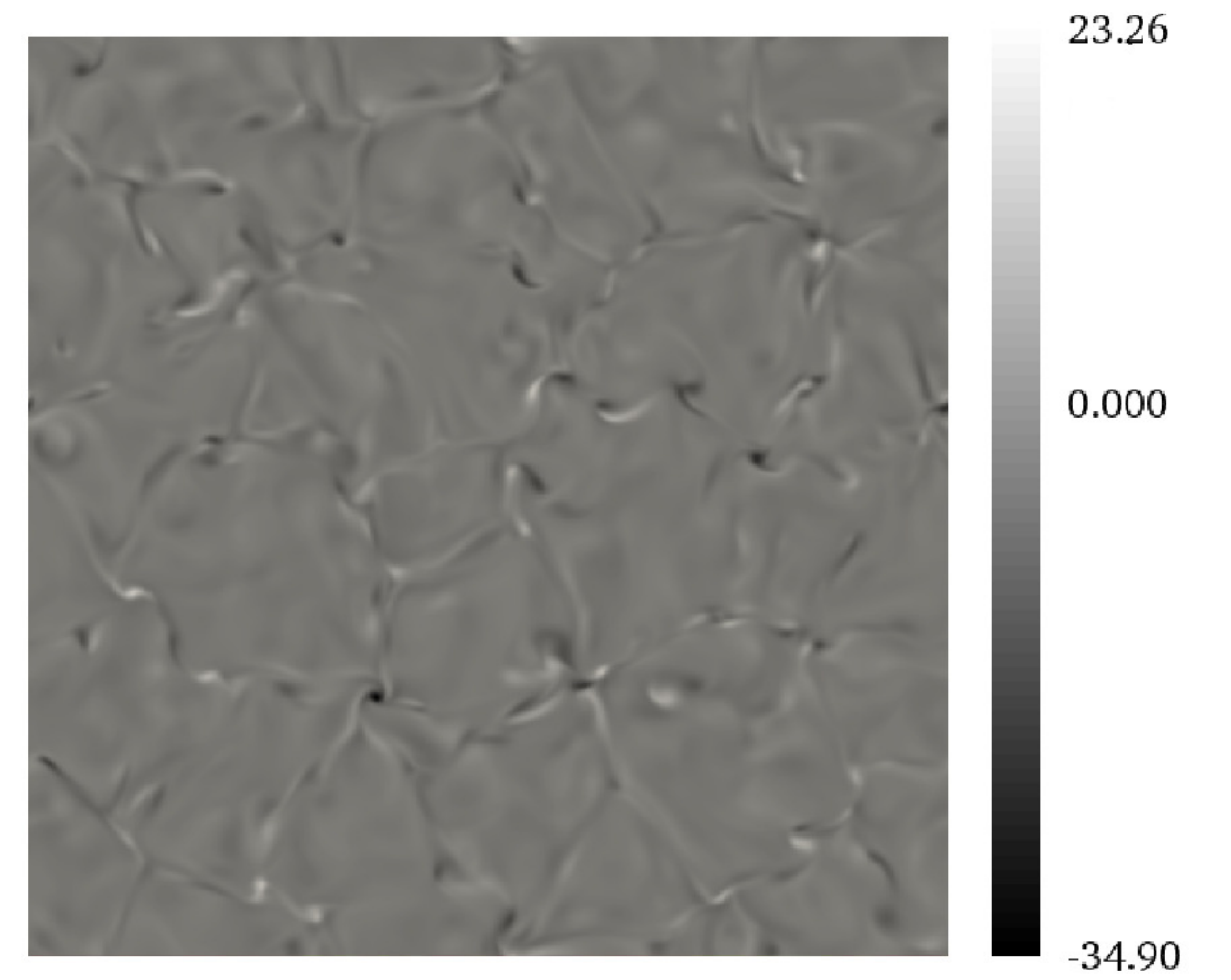}{0.45\textwidth}{}
          }
\gridline{\fig{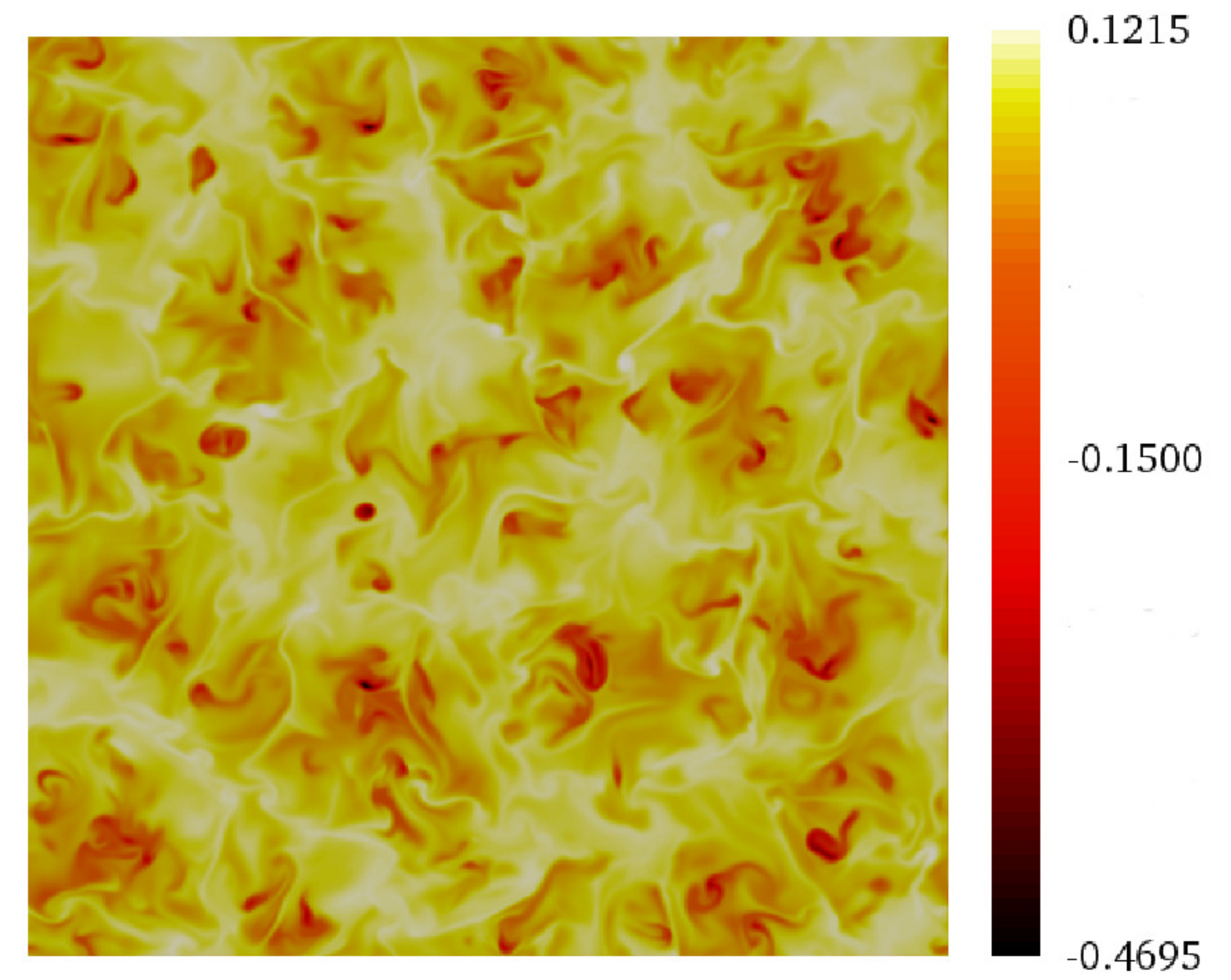}{0.45\textwidth}{(a)}
          \fig{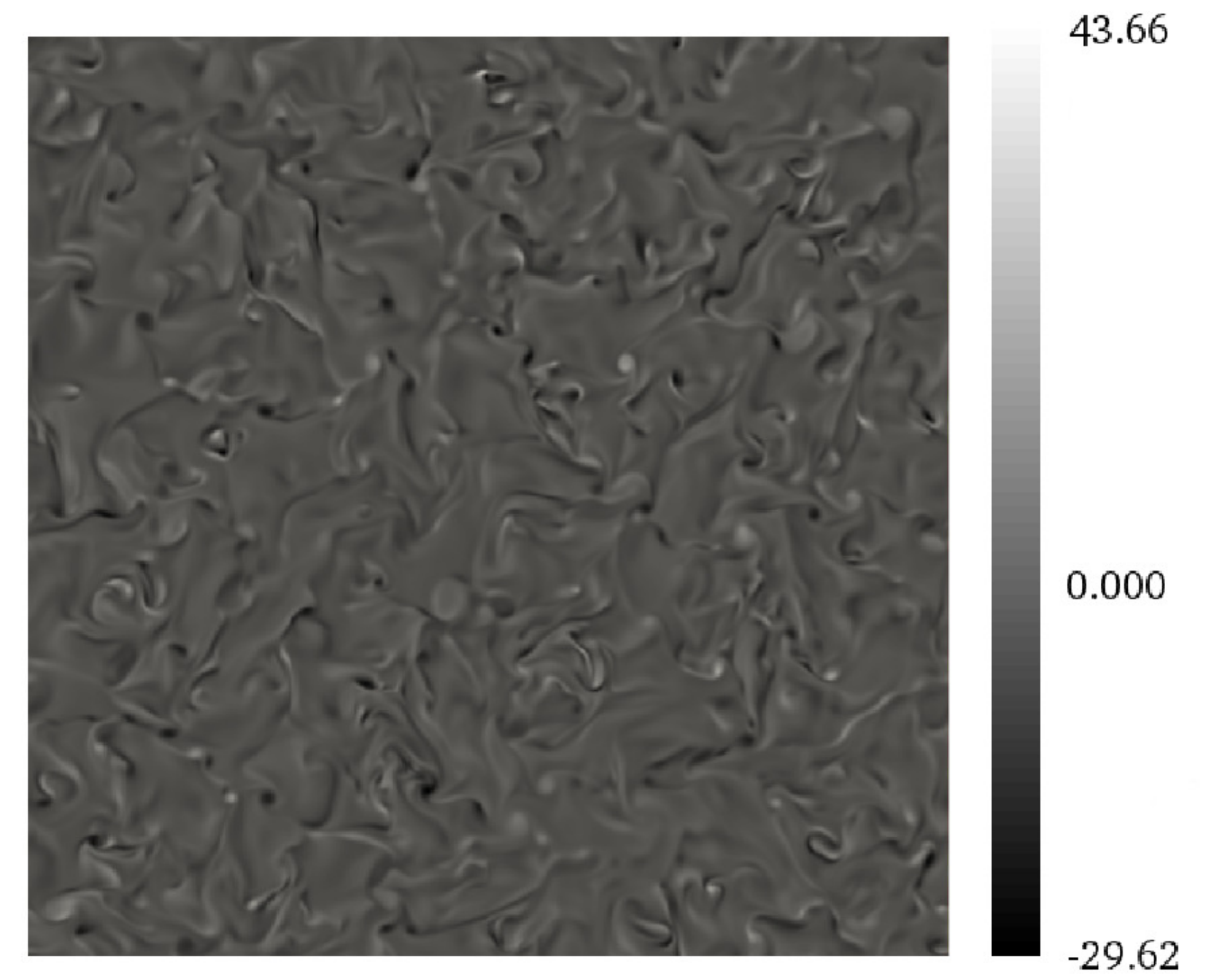}{0.45\textwidth}{(b)}
          }\caption{Horizontal slices of (a) entropy fluctuations and (b) vertical vorticity for the AC50 configuration, at heights  $z=0.9$ (top row), $z=0.5$ (middle), and $z=0.1$ (bottom).
            \label{Horizontal_slices_AC_50}}
\end{figure*}

The nature of the convective network for the strongly-stratified AC50 case can be seen in Figure~\ref{Horizontal_slices_AC_50}, which shows slices of the entropy fluctuations and the vertical vorticity at three heights in the domain. There is now a marked asymmetry between the top and bottom of the domain. Considering the entropy near the top of the domain, there is a well-defined laminar network of cells, with no evidence of turbulent small-scale behavior. Comparison with the network for the BC configurations
(Figure~\ref{Horizontal_slices_BC_AC_4_ratio_10}a) shows that the largest
convective cells observable are somewhat smaller in the strongly-stratified
case. Associated with the network in the entropy fluctuations is a
corresponding emerging network of vertical vorticity; small patches of
concentrated vorticity form where the convective cells merge. As for the the
mildly-stratified (AC4) case, it is only the cells of greatest horizontal
extent that propagate deeply; at the mid-plane, the imprint of the largest
cells at the surface survives, visible in both the entropy and vorticity. At
the bottom of the domain, the flow is turbulent, which is particularly evident
in the distribution of vertical vorticity, with no evidence remaining of the
convective network.

\subsection{Influence of the Convective Network on a Passive Scalar} \label{subsec:ink}

A complementary approach to visualising the convective network (thus sampling its Eulerian representation), and particularly to tracking its temporal evolution, is to compute the motion of passive tracer particles or `corks' (therefore, to some extent, gaining insight into the horizontal Lagrangian statistics). This idea was first introduced by \cite{Simon_1989} in order to understand the evolving photospheric network. \cite{Simon_1989} considered a model cellular flow and showed how the corks, advected solely by the horizontal component of the velocity, first moved to map out a linear network, before congregating in isolated concentrations at longer times; such behavior is reminiscent of the radial magnetic field observed in the photosphere. The notion of tracking corks has subsequently been used in numerical simulations of convectively-driven flows \citep{Cattaneo_2001, Bushby_2014_201322993}, in which the corks are again advected into the interstices of the convective network.

An alternative, though related, approach to tracking the motion of discrete passive tracer particles is to consider the temporal evolution of a continuous passive scalar --- what we shall refer to here as `ink'. In particular, we consider the advection by the horizontal velocity of an initially uniform distribution of ink. If the scalar field $\phi$ denotes the concentration of ink, then its evolution is governed by the advection-diffusion equation
\begin{equation}
\frac{\partial \phi}{\partial t} + \nabla_h \cdot \left( \vel \phi \right) = S_c(z) \nabla_h^2 \phi,
\end{equation}
where the subscript $h$ denotes horizontal derivatives. For numerical expediency, we take the Schmidt number $S_c$ to be a function of $z$, inversely proportional to $\bar \rho$ and with the value of unity at $z=0$. We are interested in the pattern mapped out by the ink on various horizontal planes; given the changes in the flow characteristics with depth, it turns out to be convenient to have $S_c$ increasing as $1/\bar\rho$.

\begin{figure*}
\gridline{\fig{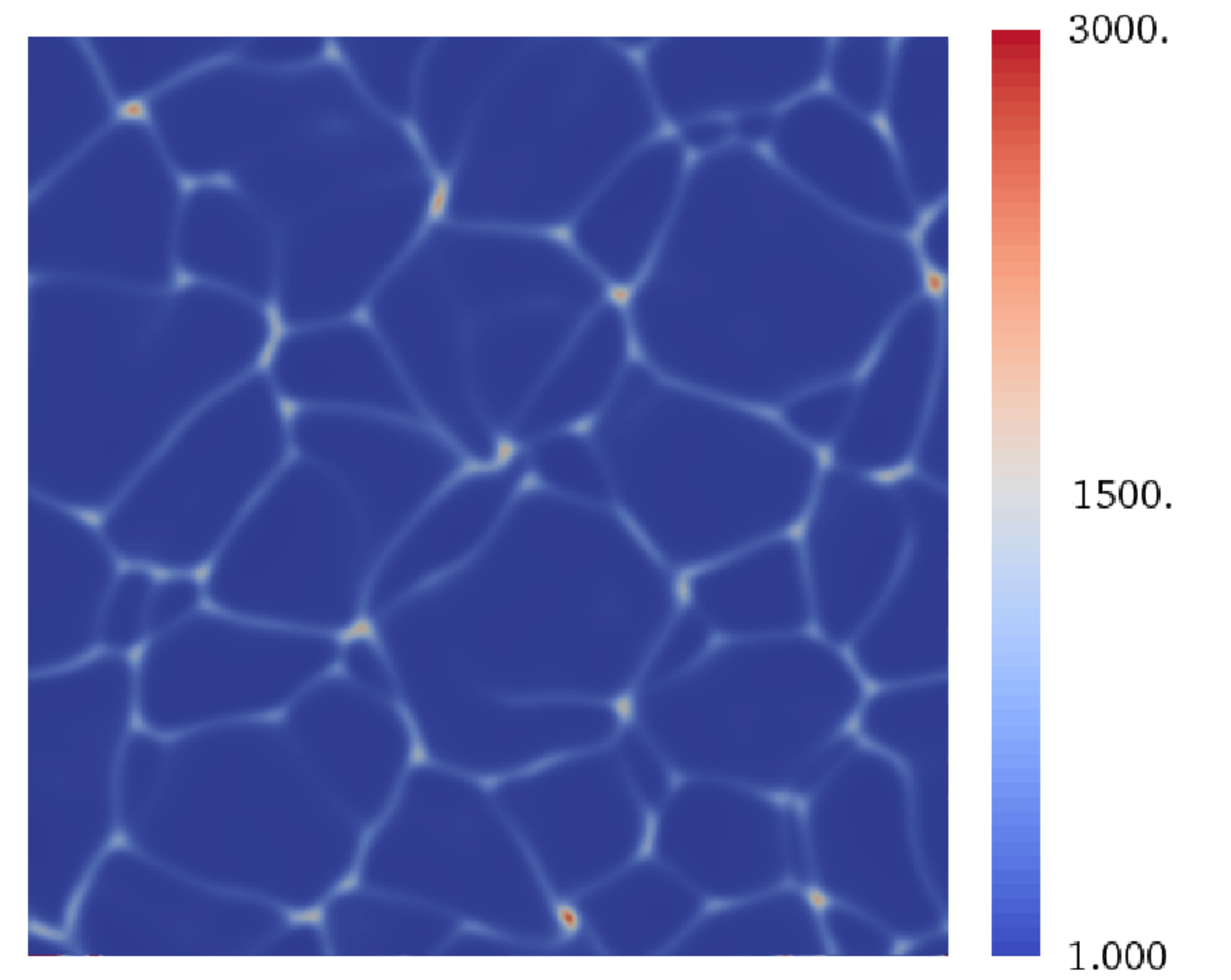}{0.45\textwidth}{}
          \fig{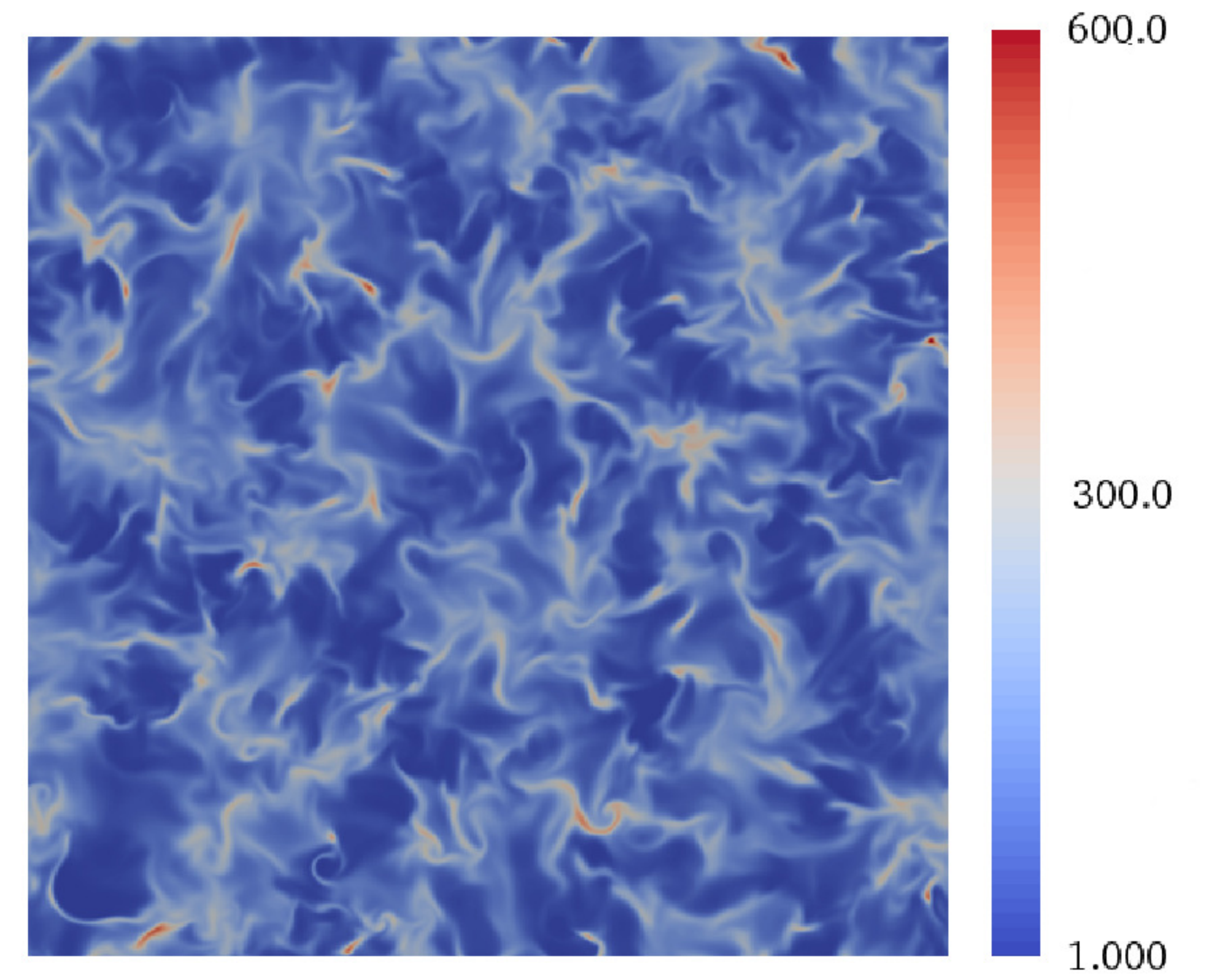}{0.45\textwidth}{}
          }

\gridline{\fig{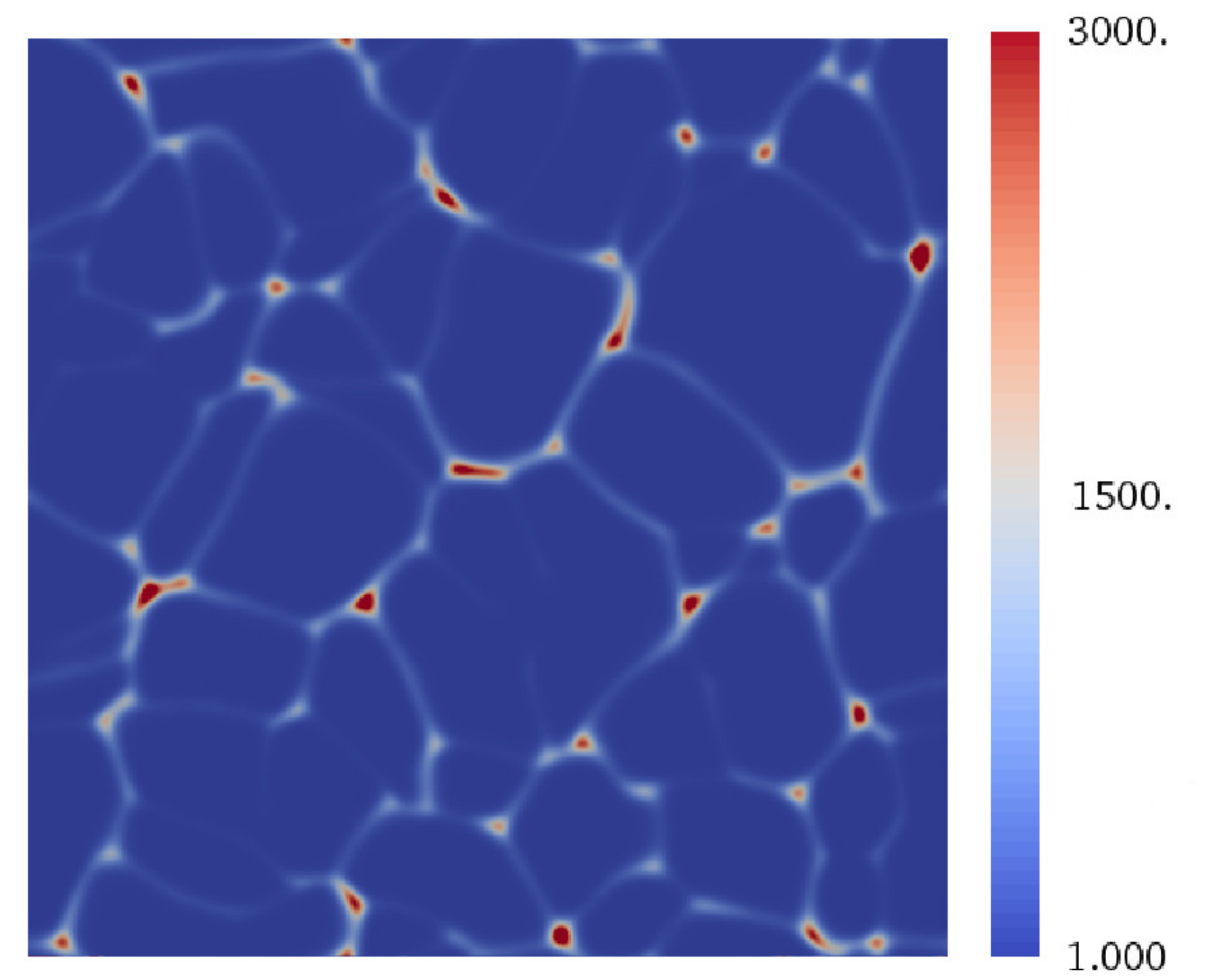}{0.45\textwidth}{}
          \fig{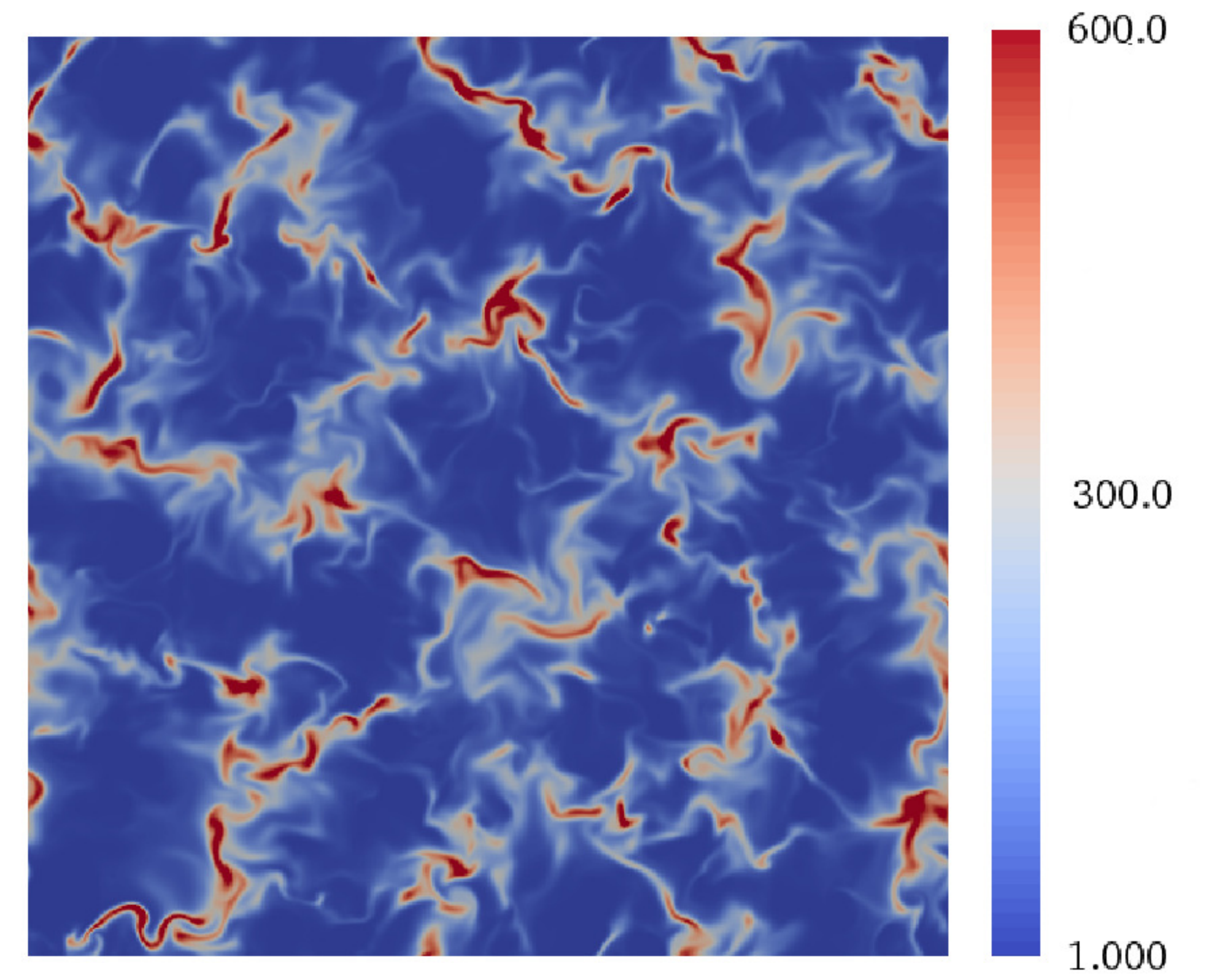}{0.45\textwidth}{}
          }

\gridline{\fig{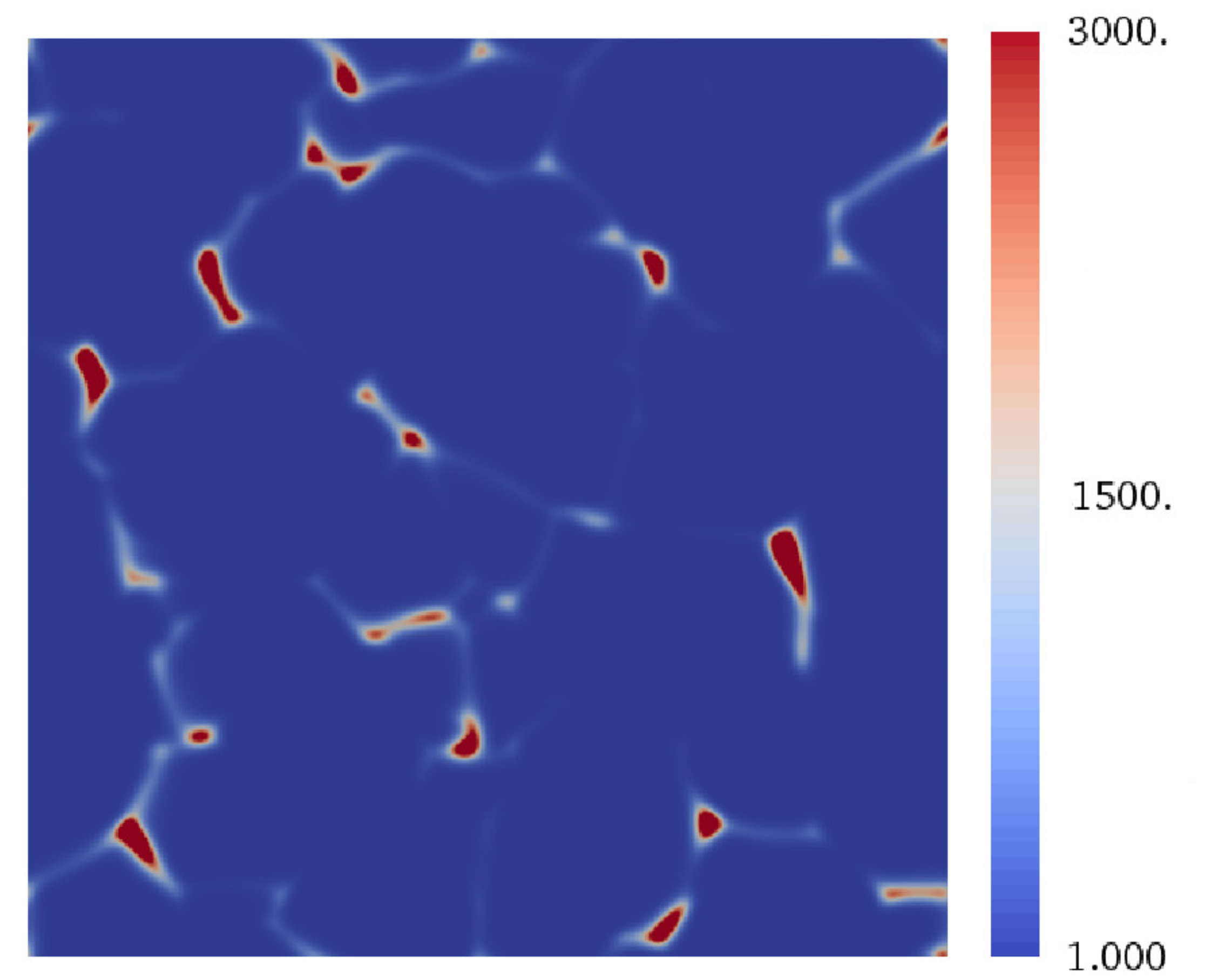}{0.45\textwidth}{(a)}
          \fig{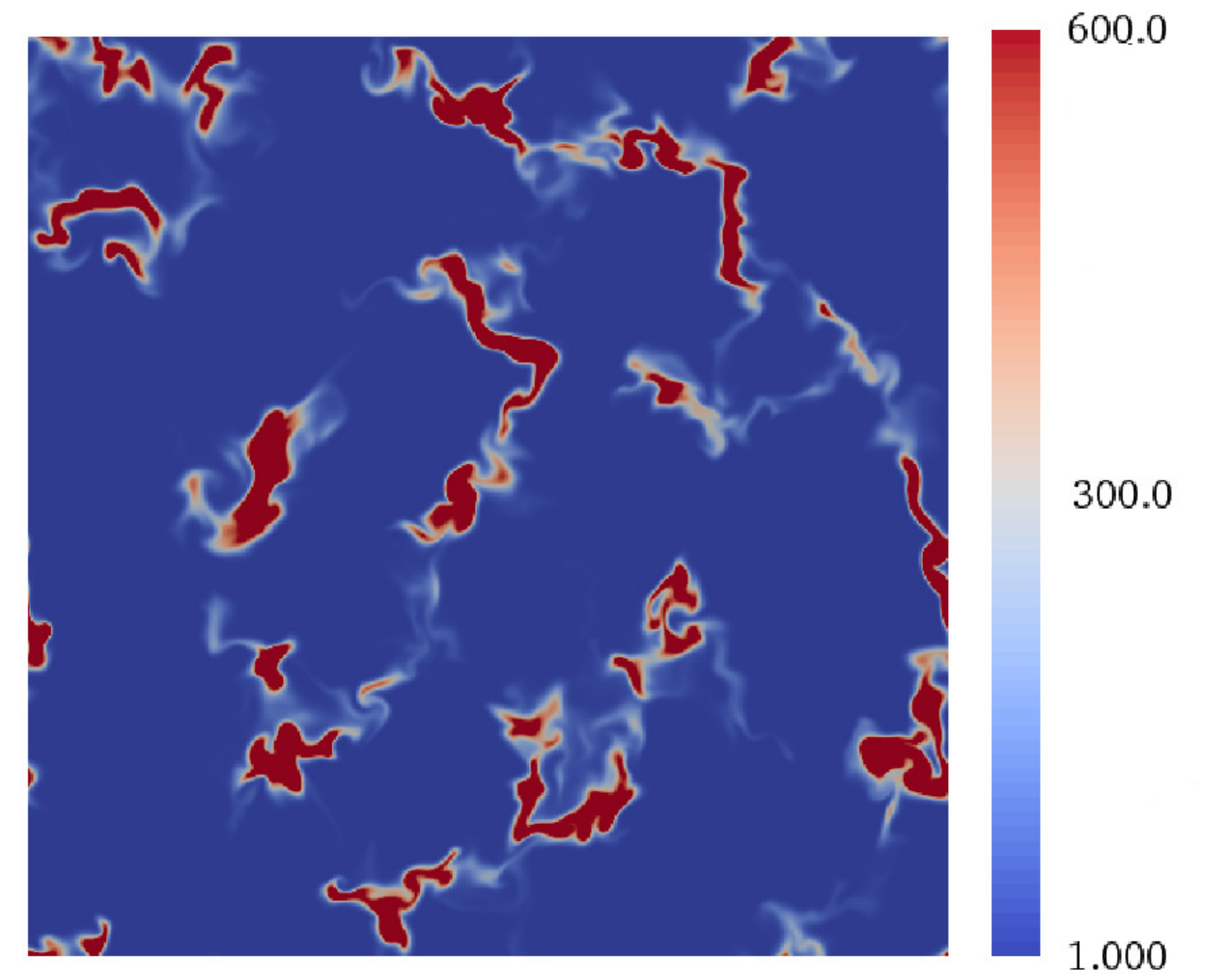}{0.45\textwidth}{(b)}
          }\caption{Distribution of ink at (a)~$z=0.9$ and (b)~$z=0.1$ at $t=0.00375$ (top), $t=0.009075$ (middle) and $t=0.0525$ (bottom). Initially the ink is uniformly distributed with a value arbitrarily fixed to $\phi = 100$. 
          The color scale is the same for all plots, chosen to cover the entire range at $t=0.00375$. The turnover time is $\tau \approx 0.025$ at $z=0.9$, and $\tau \approx 0.036$ at $z=0.1$. 
             \label{fig:ink}}
\end{figure*}

Here we consider the evolution of the passive scalar for the strongly stratified case (AC50). Figure~\ref{fig:ink} shows snapshots of the distribution of ink near the top and bottom of the domain at three representative times. At the top of the domain, the ink is pushed quickly between the cells, mapping out the convective network, as shown by the distribution at $t=0.00375$. The ink then starts to become concentrated at the corners of the cells, but with the overall network still visible, as shown by the plot at $t=0.009075$. The long-term distribution, as shown by the plot at $t=0.0525$, is marked by a disjoint concentration of ink at the corners of the convective network. The behavior at the bottom of the domain is slightly different. At $t=0.00375$, there is a hint of a network emerging, but, owing to the turbulent motions, it is much harder to distinguish than near the top. At $t=0.009075$, a clearer network is emerging at the bottom of the domain, although the turbulent motions are still visible. At long times ($t=0.0525$), the ink again accumulates in a few locations, although, in contrast to the top of the domain, it is redistributed by the turbulent motions. The most noticeable characteristics of the evolution of a continuous passive scalar  --- its accumulation between convective cells and its redistribution by the turbulent motions --- are very similar to those observed by \cite{Bushby_2012_638067} in a weakly stratified system  using discrete tracer particles.

\subsection{Spectral Distribution of Kinetic Energy}
\label{subsec:spectra}

In Sections~\ref{subsec:Convective_net} and \ref{subsec:ink} we have seen the emergence of a convective network through the temporal evolution of the entropy and of a passive scalar. Here we supplement these studies by investigating the spectral distribution of kinetic energy. In order to do so, we employ a formulation of the kinetic energy spectrum used extensively in simulations of turbulent convection \citep[e.g.][]{Bushby_2012_638067}. For a given height, $z$, and time, $t$, we evaluate the spectral distribution of kinetic energy
\begin{equation}
E_k(k_h, z, t) = \frac{\bar{\rho}(z)}{2} \sum \limits_{k_x} \sum \limits_{k_y}  |{\hat \vel} _{k_x,k_y}(z,t)|^2 ,
\end{equation}
where $\hat{\vel}_{k_x,k_y}$ is the two-dimensional Fourier transform (in horizontal planes) of the velocity field, with the summations over all horizontal wavenumbers such that $k_x^2+k_y^2 = k_h^2$. 

\begin{figure}[ht!]
\epsscale{1.20}
\plotone{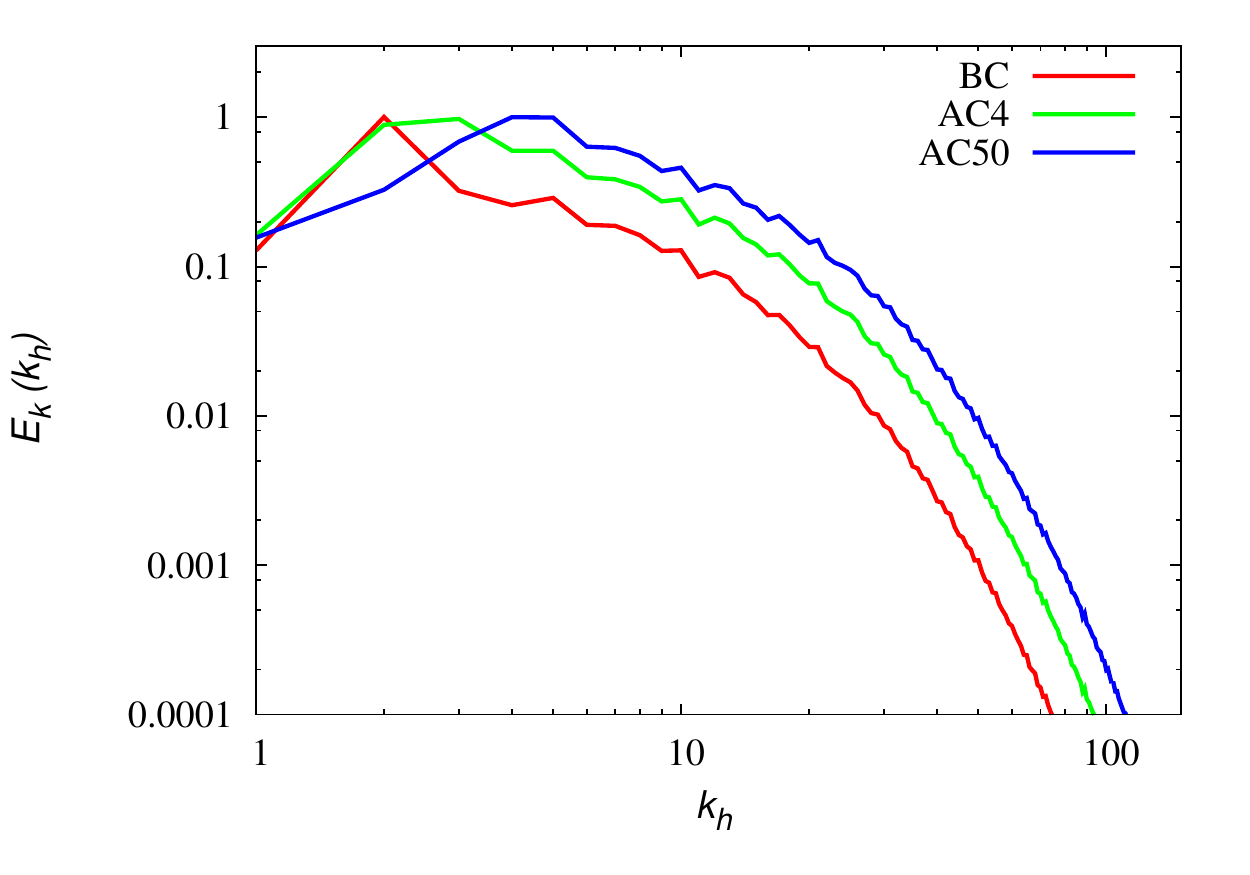}
\caption{Time- and depth-averaged kinetic energy spectra for the three different cases with aspect ratio $\lambda = 10$.
\label{fig:spectra_depth_averaged}}
\end{figure}

Figure~\ref{fig:spectra_depth_averaged} shows spectra for the three cases with $\lambda=10$ that have been both depth averaged and time averaged, over $10$--$15$ turnover times. In this figure alone, the energy spectra have been normalized so that the changes in spatial scales between the different cases become more apparent.
There is a clear peak of energy at a large scale, with the peak moving to a larger wavenumber (smaller scale) as the stratification increases. The difference between the BC and AC4 cases remains small, but is more significant in the case of AC50.

\begin{figure}
\gridline{\fig{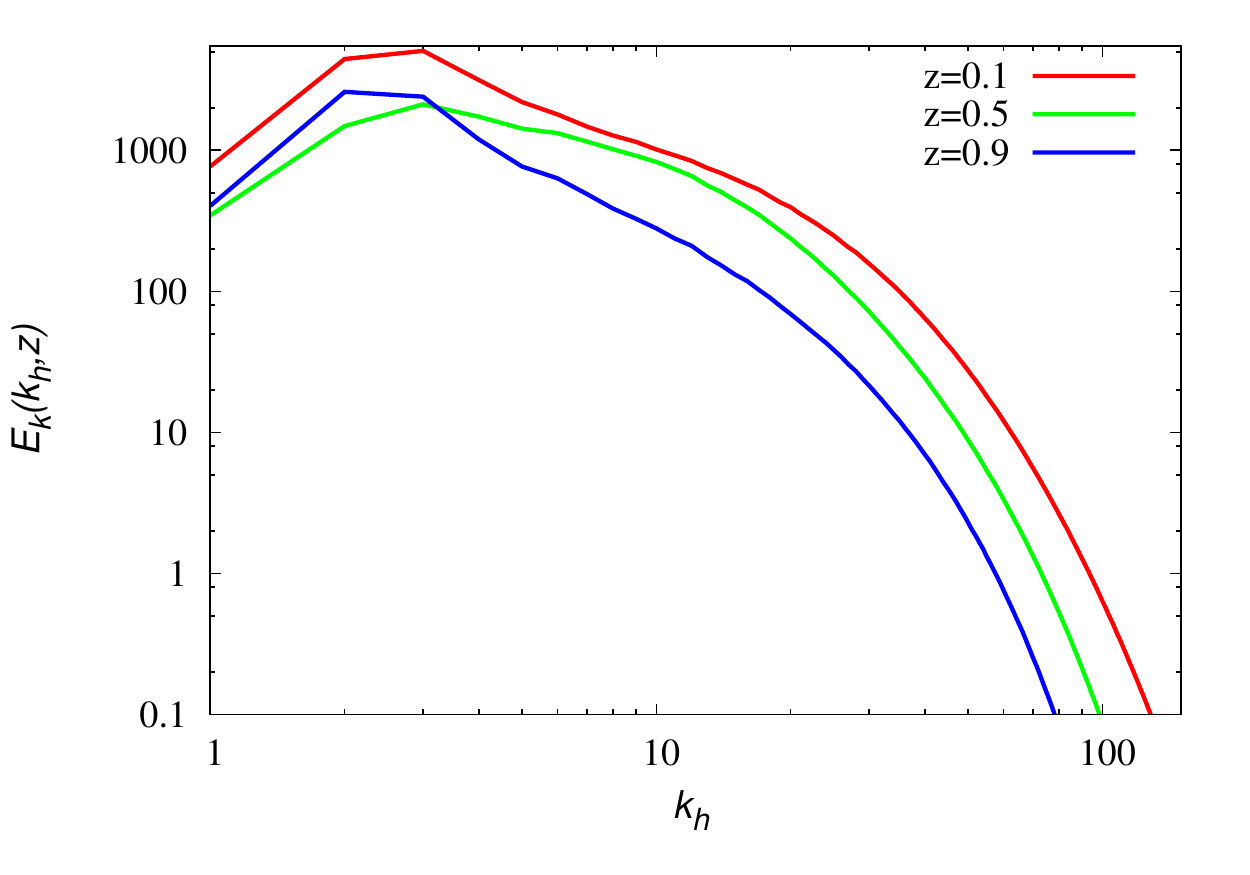}{0.5\textwidth}{(a)}}
\gridline{\fig{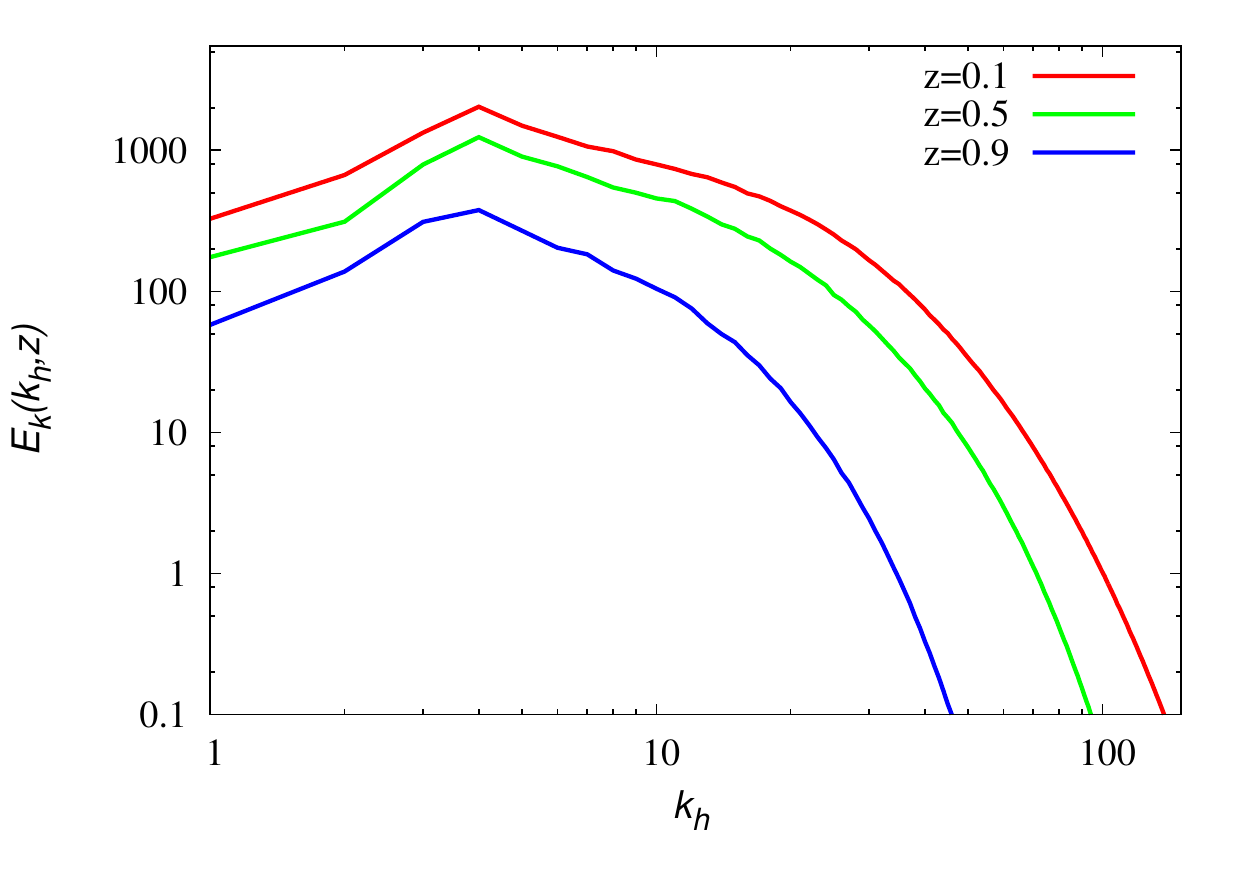}{0.5\textwidth}{(b)}}
          \caption{Kinetic energy spectra at different heights for (a)~AC4 and (b)~AC50 configurations.
   \label{fig:spectra_diff_depth}}
\end{figure}

Whereas Figure~\ref{fig:spectra_depth_averaged} provides an idea of the distribution of energy over horizontal scales for the domain as a whole, it is important also to consider the energy distribution at different heights. Figure~\ref{fig:spectra_diff_depth} shows the kinetic energy spectra at $z=0.1$, $0.5$ and $0.9$ for the AC4 and AC50 cases. Two significant observations can be made. First, the energy contained in the small scales is significantly reduced with height; this is consistent with the snapshots of the entropy and the passive scalar presented earlier and also with the studies of compressible convection by \cite{Bushby_2012_638067} and \cite{Bushby_2014_201322993}. From equations~\eqref{eq:mom} and \eqref{eq:entropy}, it can be seen that the dissipation coefficients scale as $1/\bar{\rho}$; the dissipation therefore increases with height, suppressing any small-scale turbulence near the top of the domain. The second point to note is that, even for the strongly stratified case,  the most energetic scale does not change with depth, as noted by \cite{Bushby_2012_638067} in their mildly stratified, fully compressible simulations. The results for the BC configuration are not shown here, but the spectra at $z=0.1$ and $z=0.9$  are identical, owing to the Boussinesq symmetry.

\begin{figure}[ht!]
\epsscale{1.20}
\plotone{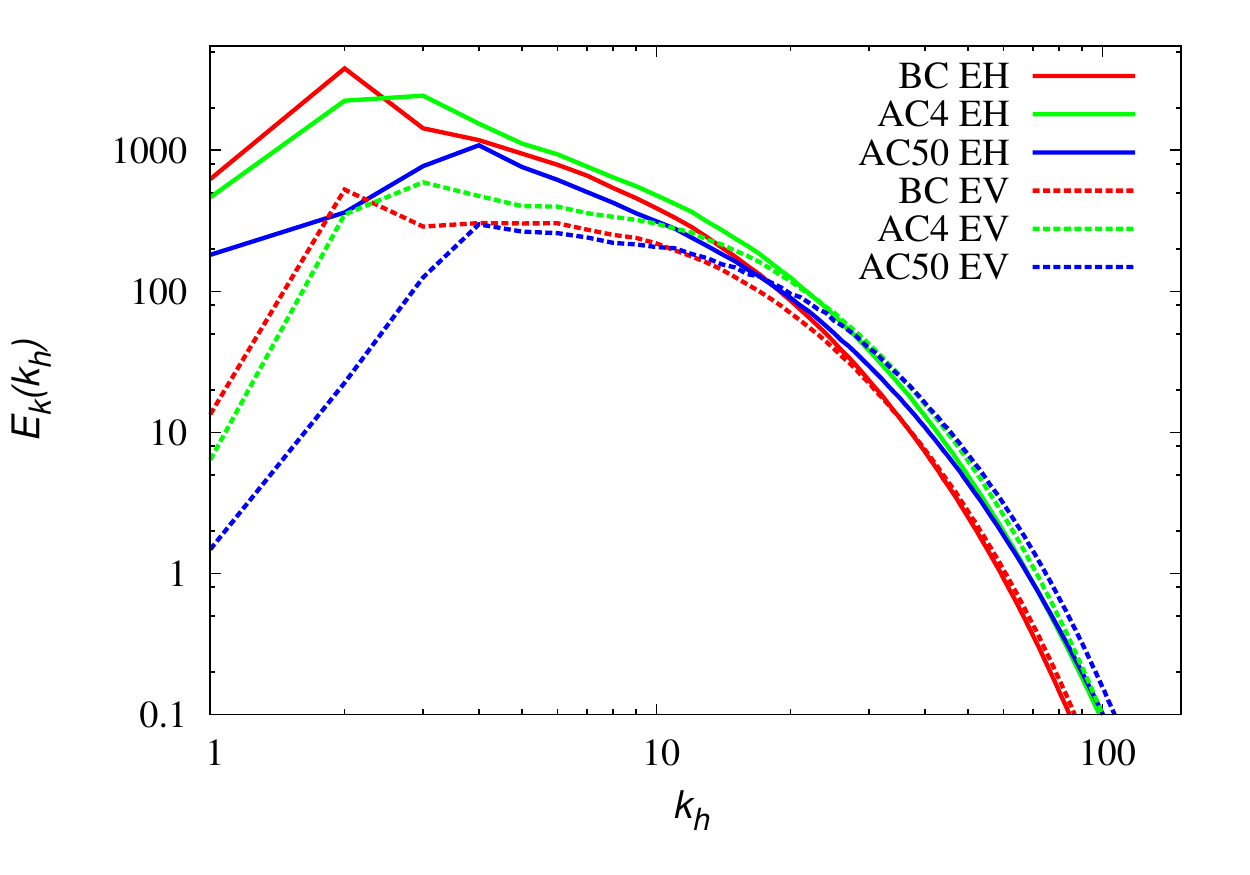}
\caption{Time- and depth-averaged spectra of the energy contained in the horizontal motions (EH) and in the vertical motions (EV), for the three different configurations with aspect ratio $\lambda = 10$.  \label{fig:Horizontal_vertical_spec}}
\end{figure}

Finally, Figure~\ref{fig:Horizontal_vertical_spec} shows, separately, time- and depth-averaged spectra of the kinetic energy associated with horizontal and vertical motions, for the three cases studied. At large scales, the bulk of the energy is contained in the horizontal motions. We note also that the migration to smaller scales with increasing stratification can be seen equally in both the horizontal and vertical spectra. It is also noteworthy that the discrepancy between the energies at large scales of the horizontal and vertical motions becomes more pronounced as the stratification increases. 
This result is reminiscent of stably stratified  turbulence \citep[see e.g.][]{Lindborg:2006,BT:2013}, where horizontal and vertical spectra are routinely calculated separately.

\newpage

\section{Discussion} \label{sec:Discussion}

In this paper we have investigated the role of stratification in modifying spatial scale selection in a local model of anelastic convection. By employing the computationally more tractable anelastic approximation, we have been able to study density ratios far in excess of those attainable in fully compressible models. Both the Eulerian and Lagrangian properties of the turbulent convection are modified by the presence of strong stratification. The midplane symmetry of the Boussinesq approximation is broken as the stratification is increased, leading to convection that has a well-defined, reasonably laminar cellular structure near the top of the domain, but a much more turbulent nature at depth. Furthermore, there is a marked shift to smaller scale convection with increasing stratification.

Here we have considered the case for which the dynamic viscosity $\mu$ and the thermal conductivity $k$ are constant, and hence the kinematic viscosity $\nu$ and thermal diffusivity $\kappa$ increase with height. An alternative prescription would be to consider the case for which $\nu$ and $\kappa$ are constant; we might then anticipate that the suppression of small-scale turbulence with height would be less pronounced.

From our simulations it is clear that in the upper regions of the domain there is a characteristic scale of convection, as shown in Figures~\ref{Horizontal_slices_BC_AC_4_ratio_10} and \ref{Horizontal_slices_AC_50}. From study of the dispersion of a passive scalar, it can be seen that the dominant scale is essentially the largest scale that emerges from the smaller-scale convective cells. One might therefore deem this to be the `mesogranular' scale. What is important to note though is that this scale cannot be attributed to a different physical formation mechanism to the smaller-scale `granular' convection. The presence or absence of this dominant ‘mesogranular’ large-scale has been attributed to the dynamics of vorticity, with strong vortex patches on small scales leading to the formation of such dominant scales \citep{Bushby_2014_201322993}. Our simulations for strong stratifications lend support to this theory and it will be interesting to determine whether changing from a constant dynamic viscosity $\mu$ to a constant kinematic viscosity $\nu$ (as discussed above) can affect the small-scale vorticity sufficiently to suppress the formation of the dominant scale. Moreover we believe the presence of rotation (as discussed below) will also modify the vorticity distribution significantly. It is also worth noting that it is very difficult to pick out spatial scales from spectral analysis since the lack of phase information puts sharp gradients in the velocity across a wide range of scales. A wavelet and multifractal analysis of the type performed by \citet{lcr:97} for solar magnetic fields may prove more revealing. 


We conclude by discussing possibilities for further investigations. At large enough scales, rotational effects will cease to be negligible; it will therefore be important to consider extended domains in order to identify how far down the spectrum the effects of rotation are felt. This will have direct relevance for understanding the interactions in the Sun between all scales from supergranules to granules. Furthermore, it is also important to identify how reducing the thermal Prandtl number to more realistic values affects scale selection and energy transport. Our study to date has been purely hydrodynamic. It will be important to extend this to consider the interaction between the convection and magnetic fields; more precisely, we shall investigate both the processing of large-scale flux generated elsewhere (thus modelling the injection of flux from a large-scale solar dynamo) and magnetic field generated by the small-scale dynamo action of the convection itself. The computationally tractable anelastic system considered here will allow the investigation of dynamo action in highly stratified turbulent domains.

\vspace{5mm}

This work was supported by STFC under grant ST/N000765/1. The computations were performed on ARC1 and ARC2, part of the High Performance Computing facilities at the University of Leeds, and on the COSMA Data Centric system at Durham University, operated by the Institute for Computational Cosmology on behalf of the STFC DiRAC HPC Facility (www.dirac.ac.uk). This equipment was funded by a BIS National E-infrastructure capital grant ST/K00042X/1, DiRAC Operations grant ST/K003267/1 and Durham University. DiRAC is part of the National E-Infrastructure.

\vspace{5mm}

\appendix
\section{Numerical Algorithm}
Here we give details of the numerical algorithm employed to solve the equations of anelastic convection (i.e.\ equations~\eqref{eq:mom}--\eqref{eq:entropy}) in Cartesian geometry. Since periodicity is assumed in the horizontal $(x,y)$ plane, we adopt standard Fast Fourier Transforms in these directions; two dimensional Fourier coefficients are indicated with hats.  The equations are discretized in $z$ using a fourth order finite-difference scheme with an evenly spaced grid. They thus form large systems of algebraic equations, which we solve via an LU decomposition using the LAPACK library.

In the anelastic formalism, $\nabla \cdot (\bar \rho \mathbf{u}) =0$
(equation~\eqref{eq:mass}); hence $\langle u_z\rangle_h = 0$ and the velocity
field can be expressed as
\begin{equation}
\bar{\rho}\bfu = \bar{\rho} \langle u_x \rangle_h \bfe_x + \bar{\rho}\langle
u_y \rangle_h \bfe_y + \nabla  \times  (\bar{\rho} \Gamma \bfe_z) + \nabla \times
\nabla \times (\bar{\rho} \tilde{P} \bfe_z ),
\end{equation}
where $\Gamma $ and $\tilde{P}$ are toroidal and poloidal scalar fields,
respectively, and the angle brackets denote horizontal averages. For simplicity in
the notation we also introduce $P=\bar{\rho} \tilde{P}$.

The momentum equation \eqref{eq:mom} can then be expressed in terms of these
toroidal and poloidal components. The evolution equation for $P$ arises from
the $z$ component of the curl of the curl of the momentum equation whilst the
evolution of $\Gamma$ is obtained from the $z$ component of the curl of the
momentum equation. This gives, in Fourier space for wavenumbers $(k_x,k_y) \ne
(0,0)$, the following equations for $\hat P$ and $\hat \Gamma$:
\begin{align}
  \label{eq:dtP}
  \frac{\partial \,\hat{ \mathrm{D}}^2 \hat P }{\partial t} & = \hat N_P + P_r \hat{ \mathrm{D}}^2 \hat{ \mathrm{D}}^2 \hat P,\\
  \text{with} \quad \hat N_P  & = -\frac{i}{k^2} \left( k_x \frac{\partial
      \hat{X}}{\partial z} + k_y \frac{\partial \hat{ Y}}{\partial z} \right)
  - \hat{Z} - R_a P_r \hat{s} + \frac{4}{3} \frac{P_r
    m^2\theta^2}{\bar{\rho}^2(1+\theta z)^2}k^2 \hat P;\\
  \label{eq:dtT}
  \frac{\partial \hat{\Gamma}}{\partial t} & = \hat N_\Gamma + \frac{P_r}{\bar{\rho}}
  \hat \nabla^2 \hat \Gamma , \qquad\text{with} \qquad \hat N_\Gamma = \frac{i}{k^2} \left( k_x \hat {Y} - k_y \hat
    {X} \right);
\end{align}
here $ k^2= k_x^2 + k_y^2$, $X= -\bfu \cdot \nabla u_x$, $Y = - \bfu
\cdot \nabla u_y$,  $Z = - \bfu \cdot \nabla u_z$ and
\begin{equation}
  \hat \nabla^2 = -k^2 + \frac{\partial^2}{\partial z^2}, \quad \hat{ \mathrm{D}}^2 = \frac{1}{\bar{\rho}}
  \left(\hat\nabla^2- \frac{m\theta}{1+\theta z}\frac{\partial}{\partial z}\right).
\end{equation}
In addition, we need to consider the evolution of the horizontally averaged
velocity. Both  non-zero components are governed by the same equation, namely
\begin{equation}
  \label{eq:dtuy}
  \frac{\partial \langle u \rangle_h}{\partial t} = 
  - \langle \mathbf{u \cdot \nabla} u \rangle_h 
  +\frac{P_r}{\bar{\rho}} \frac{\partial^2 \langle u \rangle_h}{\partial z^2},
\end{equation}
with $u=u_x$ or $u=u_y$.

These four evolution equations must be complemented by either stress-free and
impermeable boundary conditions, expressed as
\begin{equation}
  \frac{\partial \langle u_x \rangle_h}{\partial z} = 
  \frac{\partial \langle u_y \rangle_h}{\partial z} = 0 \quad\text{and}\quad
  \hat P  = \hat{ \mathrm{D}}^2 \hat P = \frac{\partial \hat \Gamma}{\partial z} = 0 \quad\text{at $z=0,1$};
\end{equation}
or no-slip and impermeable boundary conditions, which take the form
\begin{equation}
  \langle u_x \rangle_h = \langle u_y \rangle_h = 0 \quad\text{and}\quad
  \hat P = \frac{\partial \hat P}{\partial z} = \hat \Gamma  = 0 \quad\text{at $z=0,1$}.
\end{equation}
Using the same formalism, the evolution of entropy is governed, in Fourier
space, by
\begin{equation}
  \label{eq:dts}
  \frac{\partial \hat{s}}{\partial t} = \hat N_s 
  + \frac{1}{\bar \rho} \left(\hat\nabla^2 + \frac{\theta}{\bar T}
    \frac{\partial }{\partial z}\right) \hat s ,
\qquad\text{with}\qquad
  \hat N_s = \hat S + \frac{k^2 \,\hat P}{\bar \rho (1+\theta z)}  - \frac{\theta \,\hat Q}{Ra
    \,\bar\rho \bar T},
\end{equation}
where $S = -\bfu \cdot \nabla s$ and $Q$ is defined in equation \eqref{eq:Q}.
The boundary conditions for the entropy are $\hat s =0 $ at $z=0,1$.

We note that the Fourier modes $\hat \Gamma(k_x,k_y)$ and $\hat s(k_x,k_y)$, and the
averaged velocities, $\langle u_x \rangle_h$ and $\langle u_y \rangle_h$, are all
solutions to formally similar mathematical problems, which can be expressed,
for a generic field $F(t,z)$, as an evolution equation of the form
\begin{equation}
  \label{eq:dtF}
  \frac{\partial F}{\partial t} = N_F + \mathrm{L} \, F,
\end{equation}
where $N_F$ comprises nonlinear terms and $\mathrm{L}$ is a second order differential
operator in space. Boundary conditions on $F$ are either Dirichlet or Neumann
conditions, namely $F=0$ or $\partial F/\partial z=0$ at $z=0,1$.

Here we describe in detail the procedure for advancing the generic field $F$
in time. The linear and nonlinear terms in \eqref{eq:dtF} are treated
separately. The linear terms are advanced using a second order Crank-Nicolson
scheme, and the nonlinear terms by an explicit second order Adams-Bashforth
scheme (with the first time step an Euler step). The time discretization of
equations~\eqref{eq:dtF} thus yields
\begin{equation}
  \label{eq:ABCN}
  \left(1-\frac{\Delta_t}{2} \mathrm{L} \right) F^{n+1}
  =  \frac{\Delta_t}{2}\left(3 N^{n}_F -N^{n-1}_F\right) 
  +  \left(1+\frac{\Delta_t}{2} \mathrm{L} \right) F^n,
\end{equation}
where the superscript ${n}$ represents $t=n\Delta_t$, with $\Delta_t$ being a
fixed time step. The system of algebraic equations resulting from the spatial
discretization of \eqref{eq:ABCN} and of the boundary conditions is solved, at
every time step, for $F^{n+1}$. The algorithm described here applies straightforwardly to $\hat \Gamma(k_x,k_y)$,
$\hat s(k_x,k_y)$, $\langle u_x \rangle_h$ and $\langle u_y \rangle_h$.

The poloidal field, $\hat P(k_x,k_y)$, requires special attention however, owing to the form
of the linear terms in the evolution equation~\eqref{eq:dtP} and that of the boundary
conditions. The case of stress-free boundary conditions, $\hat P = \hat{\mathrm{D}}^2 \hat P = 0$ at $z=0,1$, can be treated readily in a two-step procedure partly
analogous to that used for the other fields. First, noticing
that~\eqref{eq:dtP} is an equation of the form~\eqref{eq:dtF}, for $F=\hat{\mathrm{D}}^2
\hat P$, with boundary conditions $F=0$ at $z=0,1$, we use the
algorithm~\eqref{eq:ABCN} to compute $F^{n+1}$. Then we solve, for $P^{n+1}$,
the discretized version of the elliptic equation
\begin{equation}
  \hat{ \mathrm{D}}^2 \hat P^{n+1} = F^{n+1} \quad\text{with $\hat P^{n+1}=0$ at $z=0,1$}.
\end{equation}

The implementation of no-slip boundary conditions, $\hat P = \partial \hat
P/\partial z = 0$ at $z=0,1$, is however not as straightforward and requires the use
of Green's functions and the computation of an influence matrix
\citep{Boronski:2007:PDF:1322570.1322699}. Indeed, in this case, it is not
possible to solve~\eqref{eq:dtF} for $F=\hat {\mathrm{D}}^2 \hat P$ as we have no 
information about $\hat{\mathrm{D}}^2 \hat P$ at the boundaries. Instead, we discretize equation~\eqref{eq:dtP}  in time, using again Crank-Nicolson and Adams-Bashforth algorithms, and reformulate the discretized equation. At every time step, we solve,
first
\begin{equation}
  \hat{ \mathrm{D}}^2 \hat \Lambda^{n+1}_0 = \frac{\Delta_t}{2}\left(3\hat{N}^{n}_P
    -\hat{N}^{n-1}_P\right) \quad\text{with $\hat \Lambda^{n+1}_{0}= 0$ at $z=0,1$};
\end{equation}
once $\Lambda^{n+1}_0$ known, we can compute $\hat{P}_{0}^{n+1}$ by solving
\begin{equation}
  \left(1-\frac{P_r \Delta_t}{2} \hat{ \mathrm{D}}^2 \right) \hat{P}_{0}^{n+1} = \hat
  \Lambda^{n+1}_{0} + \left(1+\frac{P_r \Delta_t}{2} \hat{ \mathrm{D}}^2\right)
  \hat{P}^n \quad\text{with $\hat P^{n+1}_{0}= 0$ at $z=0,1$}.
\end{equation}
At this stage, the poloidal component $\hat{P}_{0}^{n+1}$, the discrete solution
of the evolution equation~\eqref{eq:dtP}, satisfies only one of the two sets
of boundary conditions,  $P^{n+1}_{0}= 0$ at $z=0,1$. We need to invoke Green's functions and the
computation of an influence matrix to implement the second boundary conditions
on $\hat P$.
To achieve this, we construct, at every time step, a poloidal field consisting of $\hat P_0^{n+1}$ and a linear combination of two Green's functions,
\begin{equation}
\label{eq:PG1G2}
\hat P^{n+1}  = \hat P_{0}^{n+1} + a \,\hat{G}^{(1)} + b \,\hat{G}^{(2)}.
\end{equation}
The Green's functions $\hat{G}^{(1)}$ and $\hat{G}^{(2)}$ are computed, in preprocessing, by solving, for $G$,  the time-independent linear equations
\begin{equation}
\hat{ \mathrm{D}}^2 \hat \Lambda = 0 \quad\text{and}\quad \left(1-\frac{P_r \Delta_t}{2} \hat{ \mathrm{D}}^2\right) \hat G = \hat \Lambda.
\end{equation}
This system is solved twice: once, for $G^{(1)}$, with boundary conditions
\begin{equation}
\hat \Lambda = 1 \text{ at $z=0$ } ,\quad   \hat \Lambda = 0 \text{ at $z=1$ } \quad\text{and}\quad  G = 0 \text{ at $z=0,1$};
\end{equation}
and a second time, for $\hat G^{(2)}$, with boundary conditions 
\begin{equation}
\hat \Lambda = 0 \text{ at $z=0$ } ,\quad   \hat \Lambda = 1 \text{ at $z=1$ } \quad\text{and}\quad  G = 0 \text{ at $z=0,1$}.
\end{equation}
Clearly, the poloidal field defined by~\eqref{eq:PG1G2} satisfies  the condition $P^{n+1}= 0$ at $z=0,1$ and is the solution to the discretized evolution equation for $\hat P$,
\begin{equation}
  \left(1-\frac{P_r \Delta_t}{2} \hat{ \mathrm{D}}^2 \right) \hat{ \mathrm{D}}^2 \hat{P}^{n+1} = \frac{\Delta_t}{2}\left(3\hat{N}^{n}_P
    -\hat{N}^{n-1}_P\right)
   + \left(1+\frac{P_r \Delta_t}{2} \hat{ \mathrm{D}}^2\right)
  \hat{ \mathrm{D}}^2 \hat{P}^n.
\end{equation}
The coefficients $a$ and $b$ in the expression~\eqref{eq:PG1G2} can then be chosen such that the second set of boundary conditions can be satisfied. Thus, $\partial \hat P^{n+1}/\partial z=0$ at $z=0,1$  if $a$ and $b$ are solutions to the influence system
\begin{equation}
\begin{bmatrix}
    \partial_z \hat{G}^{(1) }(z=0) & \partial_z \hat{G}^{(2)}  (z=0) \\
    \partial_z \hat{G}^{(1)} (z=1) & \partial_z\hat{G}^{(2)} (z=1)
\end{bmatrix}
\begin{bmatrix}
    a        \\
    b        
\end{bmatrix}
= -
\begin{bmatrix}
   \partial_z \hat P_{0}^{n+1}(z=0)\\
    \partial_z \hat P_{0}^{n+1}(z=1) 
\end{bmatrix}.
\end{equation}
Flows with stress-free boundary conditions can be computed in a similar
fashion. Replacing the operator $\partial /\partial z$ by $\hat{\mathrm{D}}^2$
in the influence matrix guarantees that the condition $\hat{\mathrm{D}}^2 \hat
P^{n+1}=0$ at $z=0,1$ is satisfied. For convenience, our numerical code uses
influence matrices both for no-slip and stress-free boundary conditions. The additional computational cost of using Green's functions for stress-free boundary conditions, when not strictly required, is minimal.

When the horizontally-averaged velocity and both the toroidal and poloidal
fields have been advanced in time, the velocity can be updated using
\begin{equation}
\mathbf{u}
= 
\begin{bmatrix}
    \langle u_x \rangle_h        \\
    \langle u_y \rangle_h        \\
    0
\end{bmatrix}
+
\begin{bmatrix}
   \dfrac{\partial \Gamma}{\partial y} + \dfrac{1}{\bar\rho}\dfrac{\partial^2 P}{\partial x\partial z}\\[0.8em]
   -\dfrac{\partial \Gamma}{\partial x} + \dfrac{1}{\bar\rho}\dfrac{\partial^2 P}{\partial y\partial z} \\[0.8em]
    -\dfrac{1}{\bar\rho} \dfrac{\partial^2 P}{\partial x^2}-\dfrac{1}{\bar\rho}\dfrac{\partial^2 P}{\partial y^2}
\end{bmatrix}.
\end{equation}


The time stepping scheme for the horizontal passive scalar is different from that for the  velocity or the entropy fluctuations. Since there are no vertical motions, the equation can be expressed in Fourier space in the following form
\begin{equation}
\frac{\partial \hat{\phi}}{\partial t} + S_c(z_0) k^2 \hat{\phi}  =  \hat{N_{\phi}},
\end{equation}
with $N_{\phi}= -\nabla. \left( \mathbf{ u} \phi \right)$ and where $z_0$ is the height at which the passive scalar equation is solved.
The dissipative term is treated exactly by an integrating factor; the nonlinear term is handled by a second order Adams-Bashforth scheme (with an Euler first step). The time discretization for the passive scalar therefore yields
\begin{equation}
 \hat{\phi}^{n+1}   =  \exp\left( - S_c(z_0) k^2 \Delta_t \right) \left( \hat{\phi}^{n} +  \frac{3}{2}\Delta_t \hat N_{\phi}^{n} - \exp\left( - S_c(z_0) k^2 \Delta_t  \right)\frac{1}{2} \Delta_t \hat N_{\phi}^{n-1} \right).
\end{equation}
Note that the volume average of $\phi$ is constant over time, a fact that can be used to verify the accuracy of the code.

\bibliography{khkmt}

\end{document}